\documentclass[showpacs,superscriptaddress]{revtex4}

\usepackage{hyperref}
\usepackage{latexsym}
\usepackage{amsmath,amssymb,amsfonts}
\usepackage[dvips]{graphicx}

\begin{document}

\title{Anomalous diffusion in a symbolic model}

\author{H. V. Ribeiro}\email{hvr@dfi.uem.br}
\author{E. K. Lenzi}\email{eklenzi@dfi.uem.br}
\author{R. S. Mendes}\email{rsmendes@dfi.uem.br}
\affiliation{Departamento de F\'isica, Universidade Estadual de
Maring\'a, Av. Colombo 5790, 87020-900, Maring\'a, PR, Brazil}
\affiliation{National Institute of Science and Technology for
Complex Systems, CNPq, Rua Xavier Sigaud 150, 22290-180, Rio de
Janeiro, RJ, Brazil}
\author{P. A. Santoro}\email{psantoro@dfi.uem.br}
\affiliation{Departamento de F\'isica, Universidade Estadual de
Maring\'a, Av. Colombo 5790, 87020-900, Maring\'a, PR, Brazil}

\date{\today}

\begin{abstract}
We address this work to investigate some statistical properties of symbolic 
sequences generated by a numerical procedure in which the symbols are repeated 
following a power law probability density. In this analysis, we consider that 
the sum of $n$ symbols represents the position of a particle in  erratic 
movement. This approach revealed a rich diffusive scenario characterized by 
non-Gaussian distributions and, depending on the power 
law exponent {and also on the procedure used to build the walker}, we may have superdiffusion, {subdiffusion} or usual diffusion. Additionally, 
we use the continuous-time random walk framework to compare with the numerical 
data, finding a good agreement. Because of its simplicity and flexibility, this 
model can be a candidate to describe real systems governed by power laws 
probabilities densities.
\end{abstract}
\maketitle

\section{Introduction}
The studies of complex systems are widespread among the physical 
community\cite{Auyang,hjj,Barabasi,Boccara} and a large amount of these 
investigations deals with records of real numbers ordered in time
or in space. Based on these time series, the aim of these  works is 
to extract some features, patterns or laws which govern a given system. 
There is an extensive literature of statistical tools devoted to analyze 
time series. For instance, the detrended fluctuation analysis (DFA)\cite{Peng} 
can be used to examine the presence of correlations in the data. 
However, many of these analysis are not focused on the original data but in 
sub-series like the absolute value, the return value or the volatility 
series\cite{Mantegna}.

In particular, a time series can be converted into a symbolic sequence by using
a discrete partition in the data domain and assigning a symbol to each site 
of partition, technique that is known as symbol statistics\cite{Tang}. 
A priori, any data set can be mapped into a symbolic sequence by using a specific 
rule (see for instance Ref. \cite{Steuer}). A typical analysis performed for this
symbolic sequence is to evaluate its block entropy. This approach measures the 
amount of information contained in the block or the average information necessary
to predict subsequent symbols. This analysis was applied to a wide range of 
topics, including DNA sequences\cite{Voss}. In this context, 
Buiatti \textit{et al.}\cite{Buiatti} introduced a numerical model which 
generates long-range correlations among the symbols of a symbolic sequence, 
leading to a slow growth of the usual block entropy. 

{Motivated by this anomalous behavior in the block entropy, our main
goal is to construct a diffusive process based on these sequences. The
diffusive processes generated with these sequences are expected to be
Markovian or non-Markovian depending on the conditions imposed on
these sequences. For Markovian processes or short-term memory systems 
the mean square displacement grows linearly in time. 
On the other hand, non-Markovian processes or long-term memory systems
often display deviations of this linear behavior, being better described by
a power law on time with the exponent $\alpha$.
This is the fingerprint of the anomalous diffusion and depending on the $\alpha$ value
we may have superdiffusion ($\alpha>1$) or subdiffusion ($\alpha<1$) and
for $\alpha=1$ the usual spreading is recovered. Several physical systems exhibit 
this power law pattern. For instance, porous substrate\cite{i1}, 
diffusion of high molecular weight polyisopropylacrylamide in nanopores\cite{i2}, 
highly confined hard disk fluid mixture\cite{i3}, fluctuating particle fluxes\cite{i4}, 
diffusion on fractals\cite{i5}, ferrofluid\cite{i6}, nanoporous material\cite{i6a}, and colloids\cite{i7}.

In this context, the model proposed by Buiatti \textit{et al.} has an essential ingredient leading to anomalous diffusion:
the long-term memory present in their symbolic sequences. We will show that
a diffusive process based on these sequences lead to a rich diffusive picture,
where ballistic diffusion, superdiffusion, subdiffusion, and also usual diffusion
can emerge, depending on the model parameters and on the mode of construction
of the process. In addition, we compare our numerical results with analytic models
based on continuous-time random walk\cite{Montroll,Metzler,Zumofen,Klafter,Gorenflo,Gorenflo2,Barkai,Meerschaert,Zaburdaev,Sololov}.
}

This paper is organized as follows. In Section 2 we present and 
review some properties of the model. Section 3 is devoted to define erratic 
trajectories from the sequences as well as to investigate their diffusive 
behavior. A comparative analysis of this diffusive aspect with 
continuous-time random walk viewpoint is performed in Section 4. 
Finally, we end with a summary and some concluding comments.

\section{The model}
The {original} model\cite{Buiatti} is a numerical experiment that generates equally likely symbols which 
are repeated among the sequence following a power law probability density. 
In order to describe the model, let $\mathcal A=\{a_1,\dots,a_n\}$ be the set 
of symbols and $Q=\{Q_1, Q_2,\dots, Q_N\}$ represent a sequence where 
$Q_i\in \mathcal A$. To specify each $Q_i$, we initially select randomly one 
of the symbols of the alphabet $\mathcal A$ and repeated it $N_y$ times inside
the sequence, in such way that $Q_i=Q_{i+1}=\dots= Q_{i+N_y-1}$. The number 
$N_y$ is obtained from 
\begin{equation}
N_y=[y]+1 \quad \mbox{with} \quad y=A\left[\frac{1}{(1-\eta)^{1/(\mu-1)}}-1\right]\,,
\end{equation}
where $A>0$ and $\mu>1$ are \textbf{real} parameters, $\eta$ is a random variable uniformly 
distributed in the interval [0,1], and $[y]$ denotes the integer part of $y$.
By using this procedure, a typical symbolic sequence with $N=10$ and 
$\mathcal A=\{-1,1\}$ is 
\begin{equation}
  Q = \left\lbrace \underbrace{-1,-1,-1,}_{[y]+1=3}\underbrace{+1,+1,+1,+1,+1}_{[y]+1=5}
                     \underbrace{-1,-1,}_{[y]+1=2}
 \right\rbrace\,.
 \nonumber
\end{equation}
Since $\eta$ is a random number uniformly distributed, $y$ will be a positive random 
number. Moreover, $p(y)$ can be calculated in a straightforward manner leading to
\begin{equation}
 \label{p1}
  p(y)=(\mu-1) \frac{A^{\mu-1}}{(A+y)^\mu}\; .
\end{equation}
Therefore, the model basically consists in the repetitions of $N_y$ blocks of 
symbols with $N_y$ distributed according a power law of exponent $\mu$ in the 
asymptotic limit. Furthermore, the first and the second moments of $p(y)$ are 
given by
\begin{equation}
\langle y\rangle=\int_{0}^{\infty} \,y \,p(y)\, dy = \frac{A}{(\mu-2)}\, \; 
\mbox{(for $\mu>2$)}
\end{equation}
and
\begin{equation}
\langle y^2\rangle = \int_{0}^{\infty} \,y^2 \,p(y)\, dy  = \frac{2 A^2}{(\mu-2)(\mu-3)}\,  
\; \mbox{(for $\mu>3$).}
\end{equation}
Note that when $\mu\leq 2$ both moments diverge and when $\mu\leq 3$ the second 
moment diverges while the first one remains finite. Thus, when $\mu$ is close to 
2, $N_y$ can be very large filling a significante part of the sequence $Q$
with the same symbol. On the other hand, very large 
values of $N_y$ become rare for $\mu$ greater than 3 which makes the sequence 
highly alternating.

{It is well established that this method of building symbolic sequences generates
long-range correlations between elements of the sequence characterized by a power
law correlation function (see the analytical development of Buiatti \textit{et al.} \cite{Buiatti}). 
It was also studied that correlations lead to a non-linear growth of the usual block 
entropy, i.e., the usual entropy is not extensive for $\mu<3$. In such context, these
sequences were investigated in the framework of the so called non-extensive Tsallis 
statistical mechanics. In particular, it was shown that the 
Tsallis block entropy $S_q$ can recover extensivity for a specific choice of the 
entropic index $q$\cite{Buiatti,Ribeiro}.
}

\section{Diffusive process}
\begin{figure}[!ht]
\centering
\includegraphics[scale=0.31]{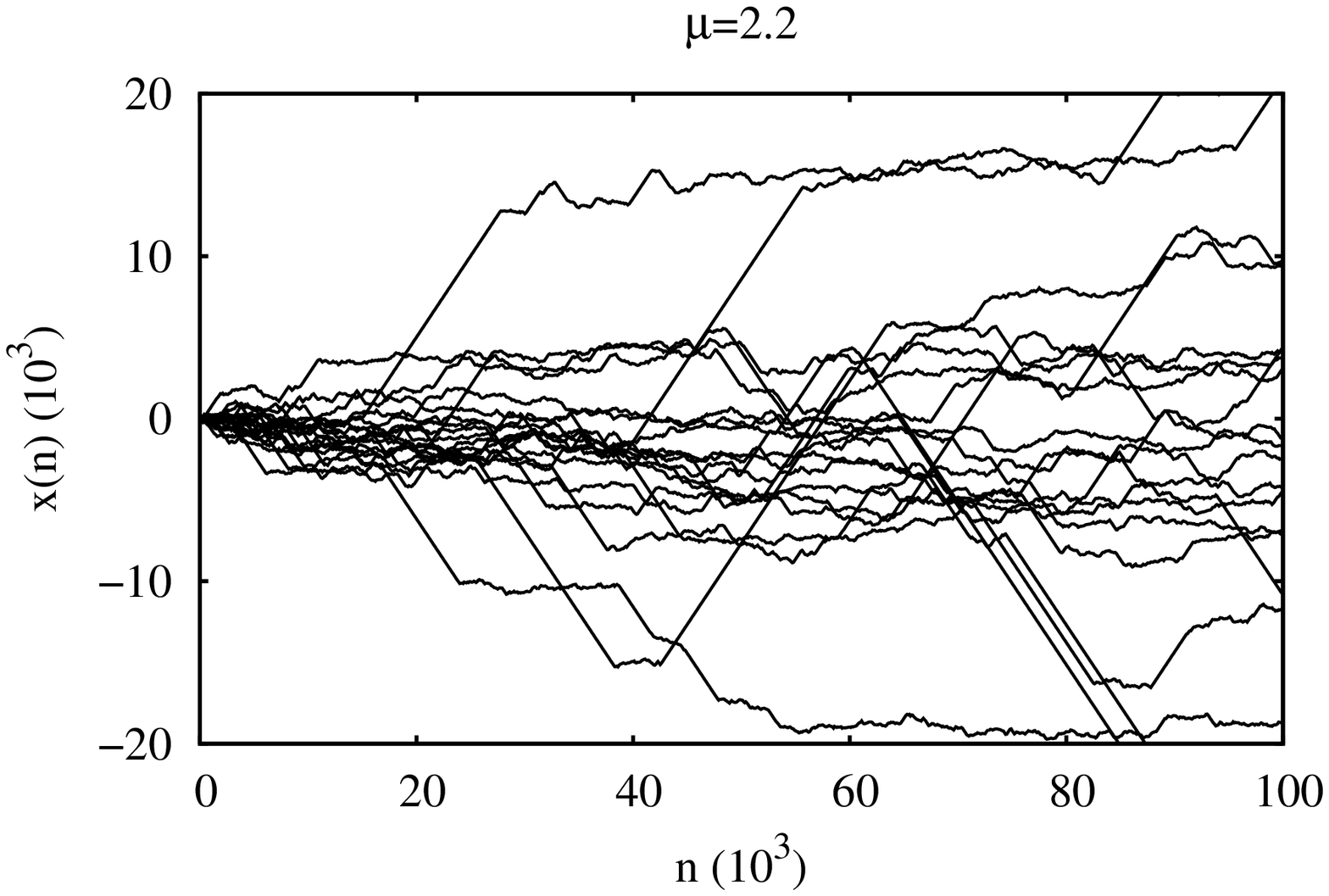}
\includegraphics[scale=0.31]{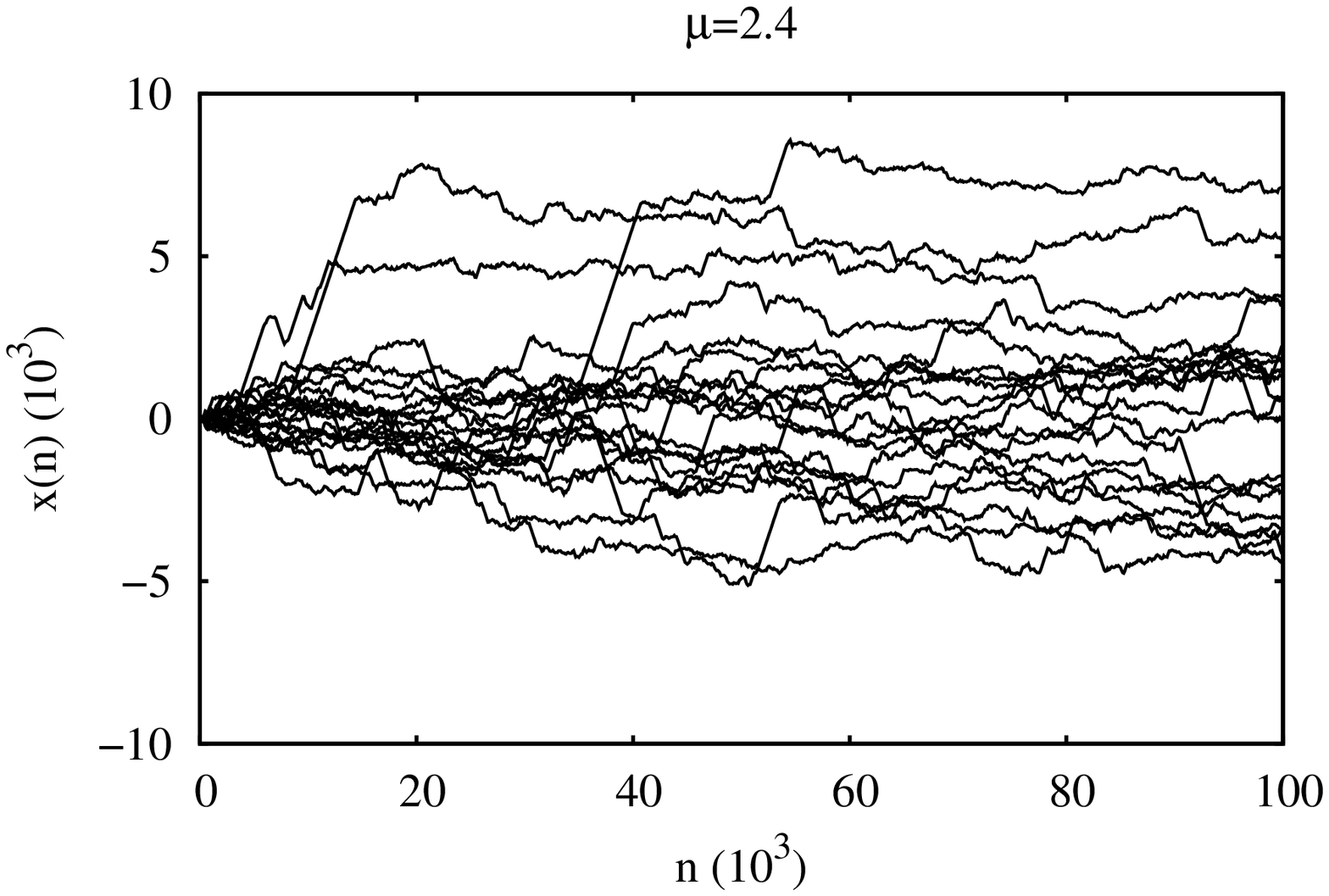}
\includegraphics[scale=0.31]{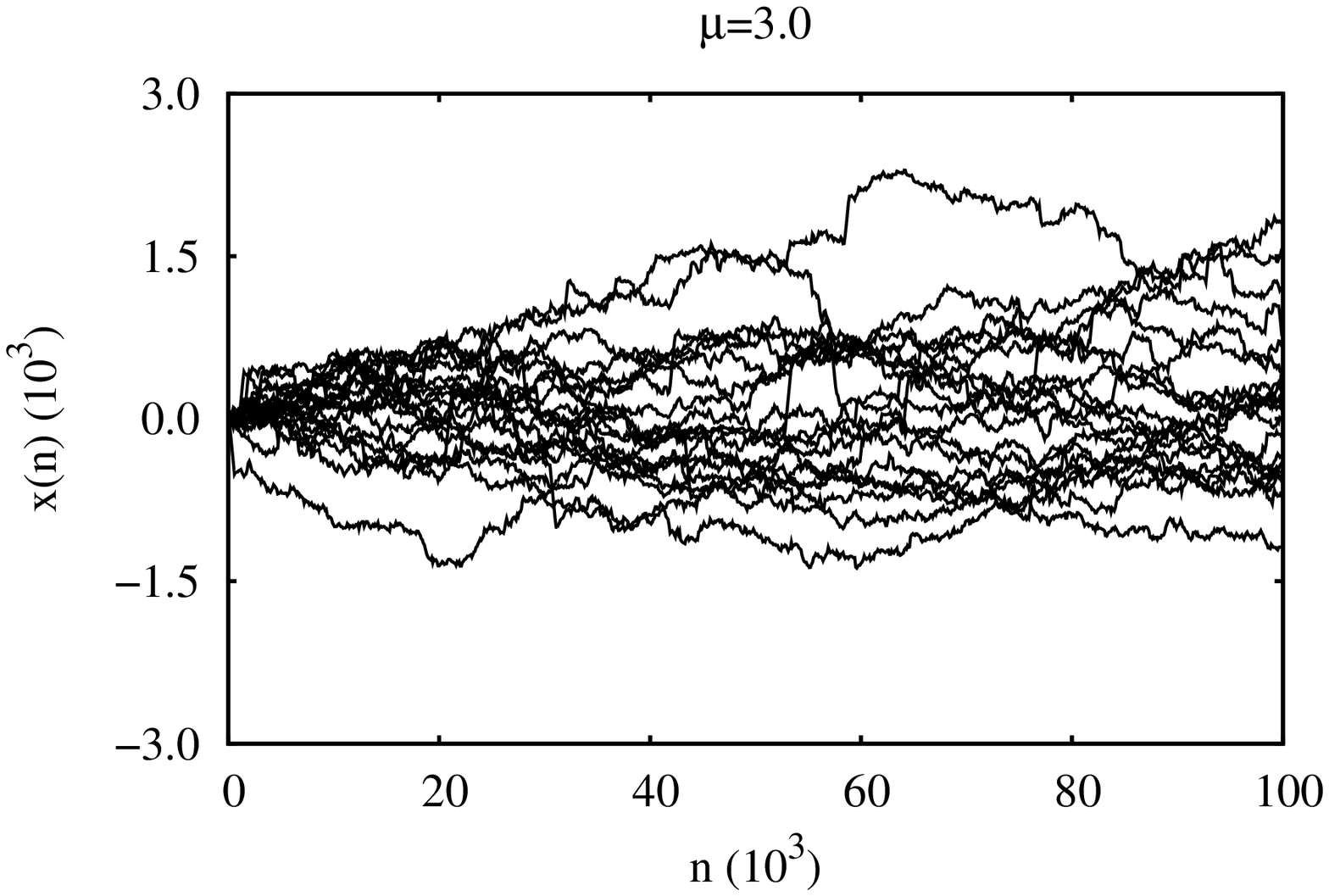}
\caption{Erratic trajectories $x(n)$ for three values of $\mu$ when considering
the two symbol alphabet $\mathcal A=\{-1,1\}$ and $A=1$. Note that
depending on the values of $\mu$ the trajectories are very different.}
\label{fig:traj}
\end{figure}

{As we raised in the introduction, long-term correlations or non-Markovian 
processes frequently  present anomalous properties when investigated in the 
context of diffusion. In this direction, to construct
erratic trajectories from these sequences may yield a rich diffusive scenario based on
a simple model. This point has been noted by Buiatti \textit{et al.} and also briefly discussed by us
in Ref. \cite{Ribeiro}. 

A simple and direct way to obtain the trajectories is to consider that each symbol in a 
sequence represents the length of the jump of a particle in erratic movement. The
position of the symbol plays the role of time. In this manner, the variable 
\begin{equation}\label{eq:pe}
 x(n) = \sum_{i=1}^{n} Q_i\,
\end{equation}
represents the position of the particle after a time $n$, which is an integer because of 
the construction.

Let us start our investigation by considering the simplest symmetric case, i.e.,
the two symbols alphabet $\mathcal A=\{-1,1\}$. Thus, the particle is equally likely to
jump to the right or to the left according to whether $Q_i = 1$ or $Q_i = -1$ and
the variable $x(n)$ represents a random walk-like process where $x(1)=\pm1$, 
$x(2)=0,\pm2$, $x(3)=\pm3,\pm1$ and so on, with $n$ playing the role of \textit{time}. 
Figure~\ref{fig:traj} illustrates $x(n)$ for three values of $\mu$ with $A=1$. 
Note that the trajectories are remarkably different depending on the values 
of $\mu$. For small values of $\mu$ ($\mu<3$), we can see that the trajectories 
are governed by two mechanisms: spatial localization and large jumps. 
When $\mu\leq 2$, larger is the jump, reflecting the fact that all 
moments of $p(y)$ diverge for $\mu \leq 2$. On the other hand,  the 
second moment of $p(y)$ is finite for $\mu>3$ and the trajectories are very 
similar to usual random walks.
}

As pointed out in the introduction, when dealing with diffusive process, it is very common to investigate how the 
particles are spreading by evaluating the variance 
$\mbox{$\sigma^2(n) = \langle(x(n) - \langle x(n)\rangle)^2\rangle$}$, where 
the angle brackets denote an ensemble average.  The usual Brownian motion\cite{Gardiner,Vlahos}  
is characterized by $\sigma^2(n) \sim n$ and by a Gaussian propagator 
$p(x,n)\sim n^{-1/2} \exp(-x^2/2 n)$ which is a direct consequence of the 
central limit theorem and the Markovian nature of the underlying stochastic 
process. On the other hand, the anomalous diffusion behavior is usually distinguished by the value of the 
exponent $\alpha$\cite{Dybiec} in 
\begin{equation}\label{eq:anomalous}
 \sigma^2(n) \propto n^\alpha\,.
\end{equation}
We have subdiffusion when $0<\alpha<1$ and superdiffusion when $\alpha>1$. 
The crossover between subdiffusion and superdiffusion corresponds
to the usual Brownian motion and the case $\alpha=2$ is the ballistic regime. 

\begin{figure}[!ht]
\centering
\includegraphics[scale=0.4]{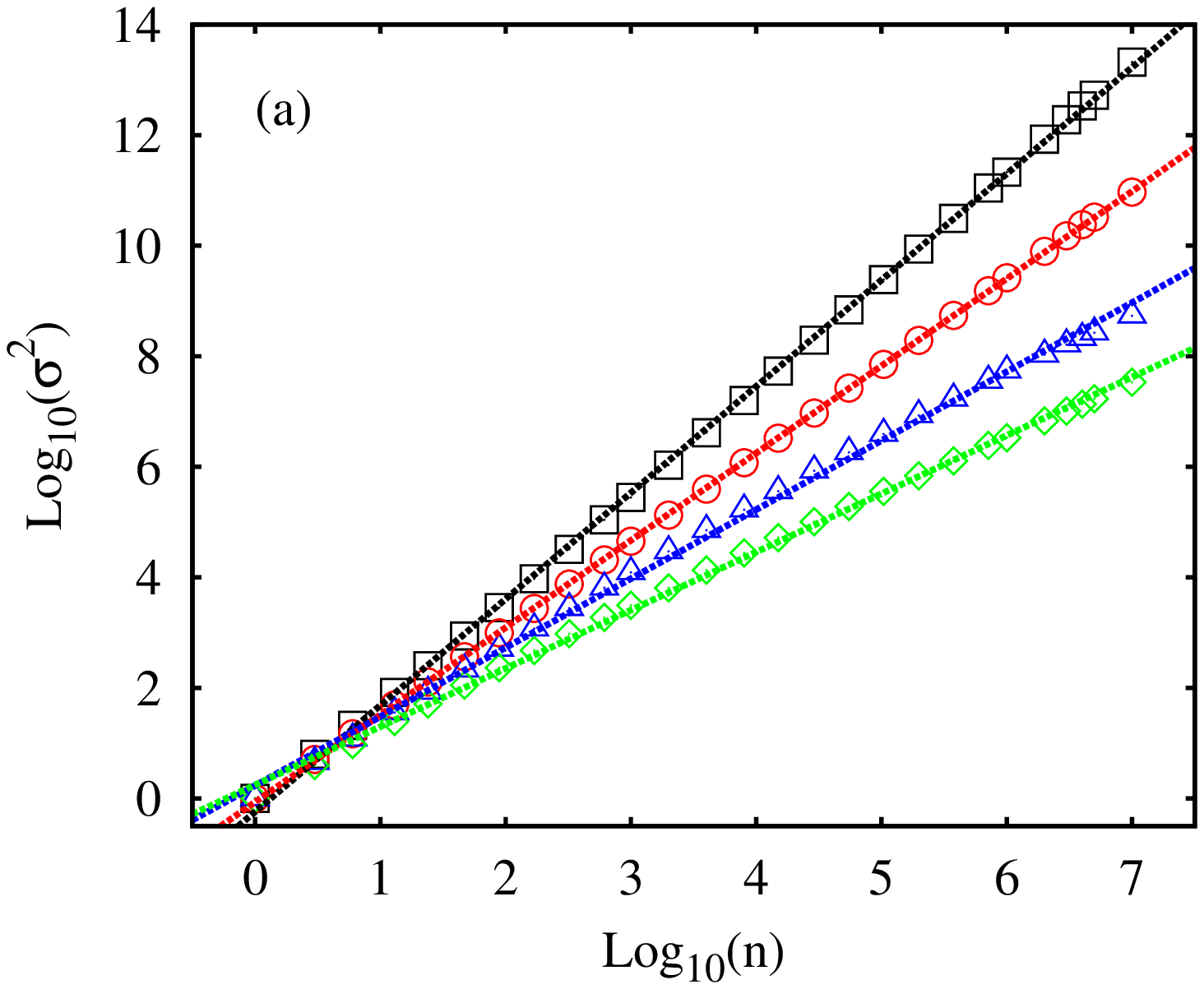}
\includegraphics[scale=0.4]{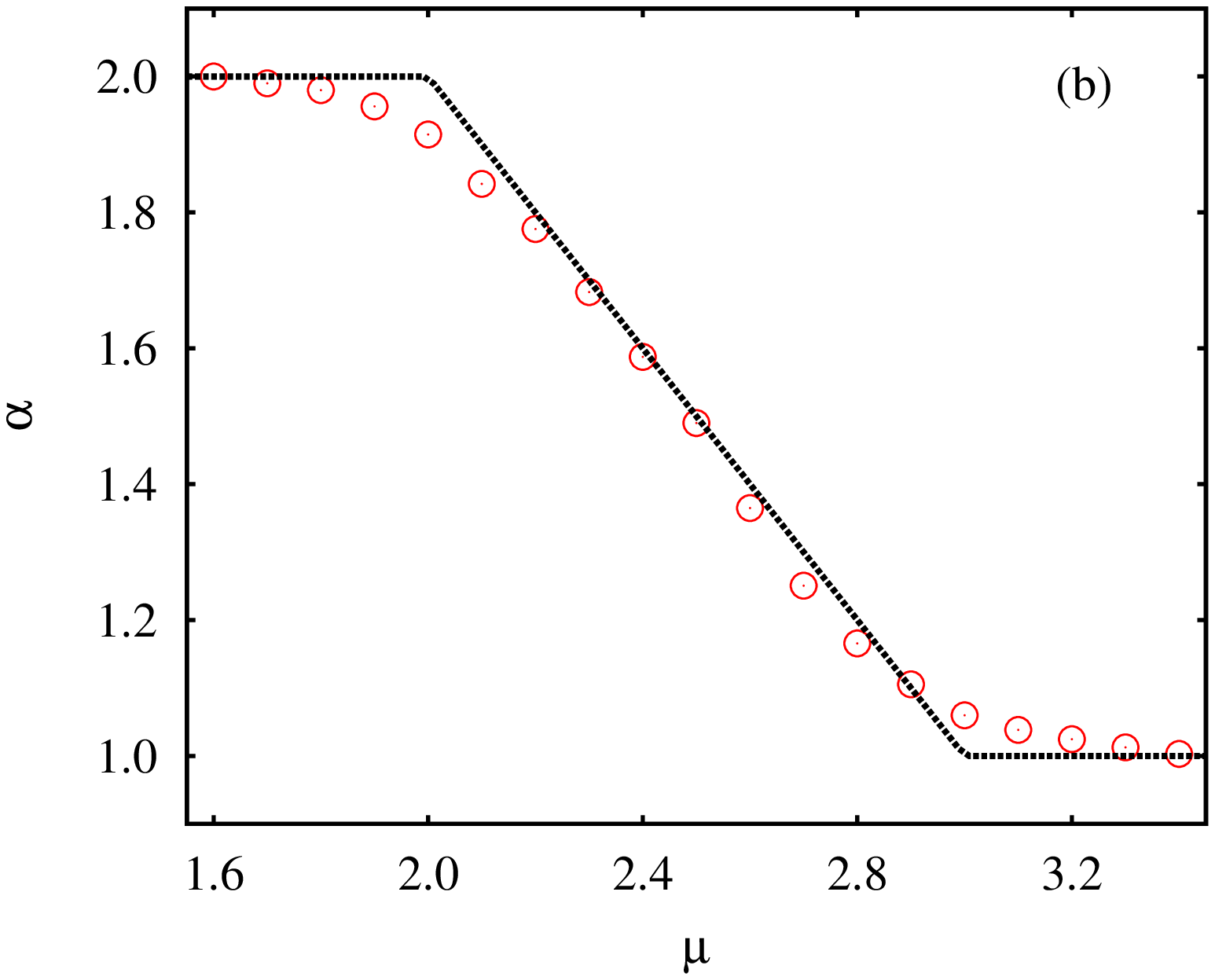}
\caption{{Results concerning the alphabet $\mathcal A=\{-1,1\}$ for $A=1$.}
 (a) Logarithm of the variance $\sigma^2(n)$ versus logarithm of $n$ for 
$\mu=1.8$ (squares), $\mu=2.4$ (circles), $\mu=2.8$ (triangles) and $\mu=3.4$ 
(diamonds). In this figure, the slopes of the curves are numerically equal to the 
exponents $\alpha$ and the straight lines are linear fit to the data used to 
obtain the values of $\alpha$. (b) The dependence of the exponent $\alpha$ on 
$\mu$. From this figure, we have basically three diffusion regimes: a ballistic 
($\mu\lesssim2$), a superdiffusive ($2\lesssim\mu\lesssim3$) and the usual 
Brownian motion ($\mu\gtrsim3$). The straight lines are the predictions of the 
continuous-time random walk model related to equation (\ref{eq:var}).}
\label{fig:var}
\end{figure}

In this direction, we evaluate the variance for several values of $\mu$ over 
an ensemble average of $5\times10^5$ realizations as shown in 
Figure~\ref{fig:var}a. In a log-log plot the slope of the curve $\sigma^2(n)$ 
versus $n$ is numerically equal to the exponent $\alpha$ which is visibly 
changing with the parameter $\mu$. In Figure~\ref{fig:var}b, we quantify this 
dependence by plotting $\alpha$ versus $\mu$. From this figure, we have basically 
three diffusion regimes depending on the existence of the first 
$\langle y \rangle$ and the second $\langle y^2 \rangle$ moments of $p(y)$: 
(i) a ballistic one for $\mu<2$ ($\langle y \rangle \to \infty$ and 
$\langle y^2 \rangle \to \infty$),
(ii) a superdiffusive one for $2<\mu<3$ ($\langle y \rangle$ finite 
and $\langle y^2 \rangle \to \infty$) and
(iii) the usual Brownian motion for $\mu>3$ ($\langle y \rangle$ and 
$\langle y^2 \rangle$ finite).

\begin{figure}[!ht]
\centering
\includegraphics[scale=0.4]{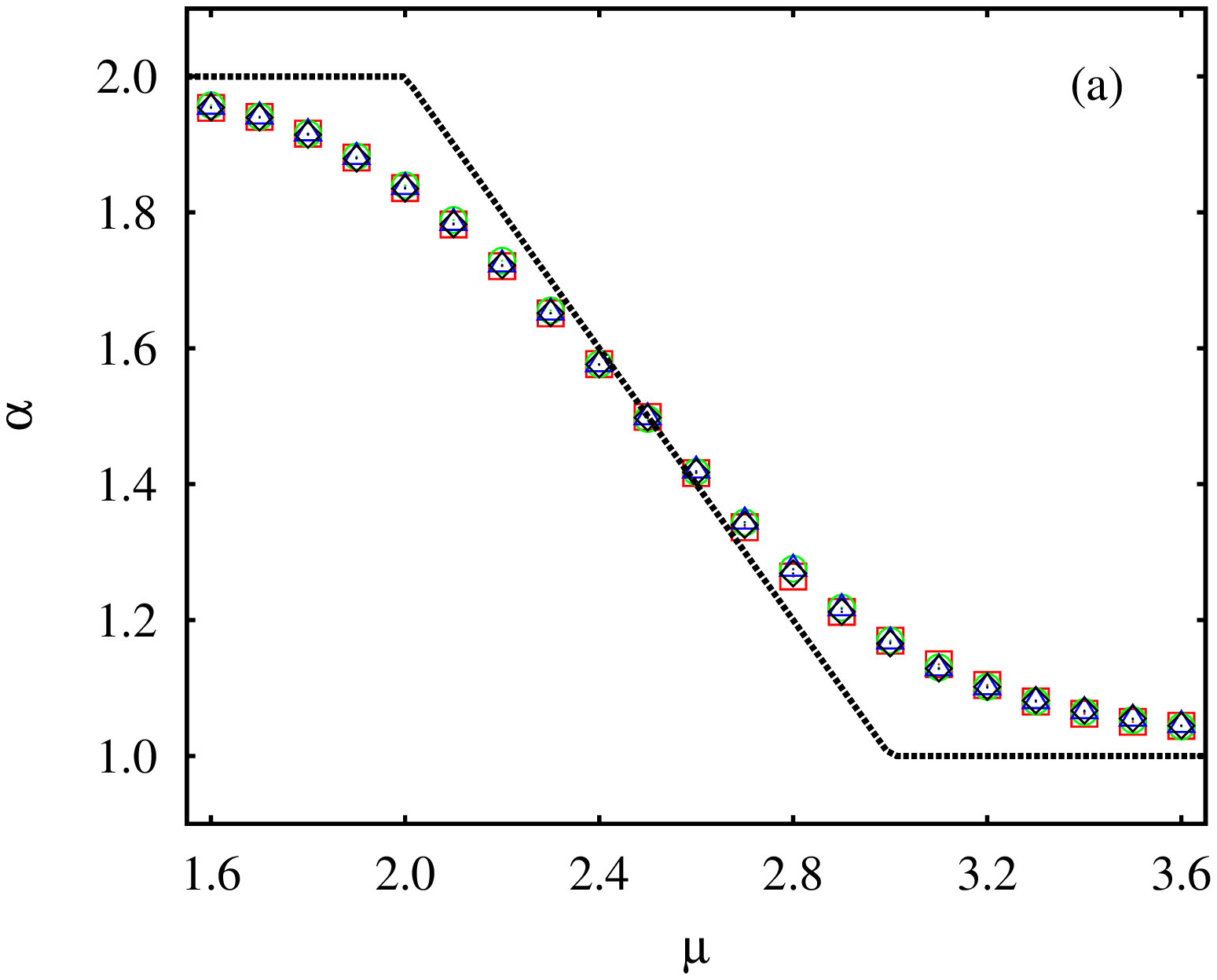}
\includegraphics[scale=0.4]{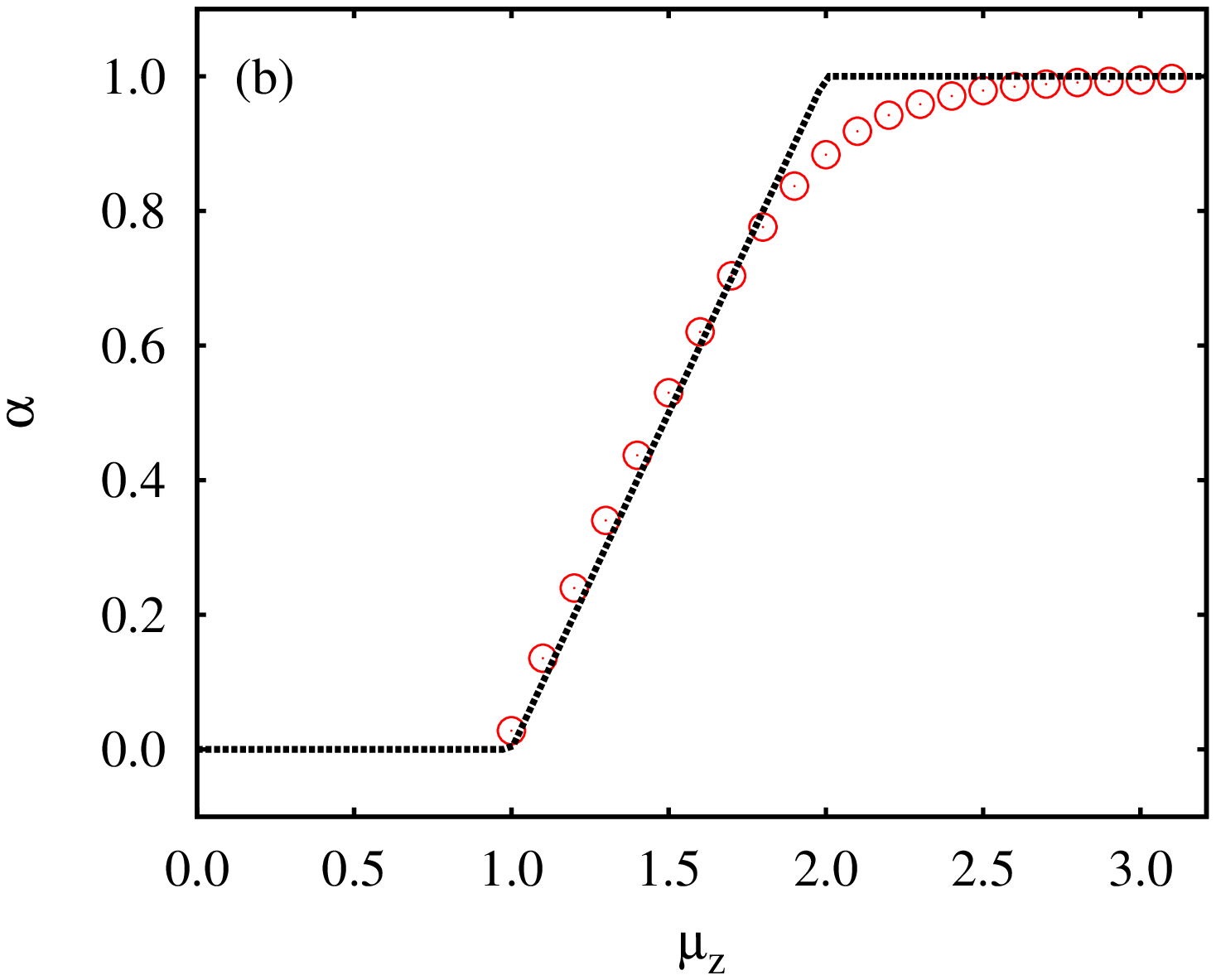}
\caption{(a) The dependence of the exponent $\alpha$ on $\mu$ when considering the alphabet 
$\mathcal A=\{-1,0,1\}$ and varying the probability of the zero symbol within the sequence.
For the squares the symbols are equiprobable and for the circles the zero symbol is twenty times more likely 
to occur ($\mathcal A=\{-1,0^{20},1\}$). The triangles and the diamonds are the results concerning the larger alphabets 
$\mathcal A = \{-2,-1,0,1,2\}$ and $\mathcal A = \{-20,\dots,-2,-1,0,1,2,\dots,20\}$, respectively.
(b) The dependence of the exponent $\alpha$ on $\mu_z$ when considering the equiprobable alphabet 
$\mathcal A=\{-1,0,1\}$ and $\mu_j=6$. Note the presence of a subdiffusive regime for $1\lesssim\mu_z\lesssim2$.
All results were obtained by using sequences of length $10^7$ averaged over $5\times10^5$ realizations
with $A=1$. The straight lines are the predictions of the continuous-time random walk model related to equation (\ref{eq:var})
and (\ref{eq:var2}), respectively.
}
\label{fig:var2}
\end{figure}
{
Next, we investigate the role of size of the symbol space by considering
that more symbols are present in the alphabet $\mathcal A$. We consider first the presence
of zeros, i.e., $\mathcal A=\{-1,0,1\}$ where each symbol is equiprobable within the sequence.
The zero symbol allows the particles to stay motionless for a certain time what could be related
to subdiffusion. However, as we show in Figure~\ref{fig:var2}a the presence of zeros in the 
sequence does not significantly change the profile of the relation $\mu$ versus $\alpha$. Moreover,
even if zero symbol becoming more probable within the sequence, i.e., $\mathcal A=\{-1,0^{T},1\}$
where $T$ is number of zero symbols in the alphabet, this result remains valid, as we also
show in Figure~\ref{fig:var2}a. Second, we study larger alphabets from $\mathcal A = \{-2,-1,0,1,2\}$ to $\mathcal A = \{-20,\dots,-2,-1,0,1,2,\dots,20\}$ and the results are shown in Figure~\ref{fig:var2}a. 
We found that the relation $\mu$ versus $\alpha$ does not depend on
the size of the symbol space. This relation is also robust for variations of the parameter $A$ and for non-symmetric
alphabets. In particular, the parameter $A$ only produces a multiplicative effect in the equation (\ref{eq:anomalous}) and a 
non-symmetric alphabet produces a drift which does not affect the spreading of the system.

Until now we were not able to generate subdiffusion, even adding the zero symbol more likely 
to occur. This result suggests that only superdiffusion can emerge when considering the same
value of $\mu$ for the symbols that lead to jumps and for the zero symbol that lead to absence 
of motion. The reasons for this behavior are related to the fact that even the zero symbol being 
much more likely than other symbols, the number of repetitions $N_y$ is independent of 
the symbol. Thus, the particles can remain at rest for a long time but the flights can be 
equally long, since for $\mu<3$ there is no characteristic scale for $N_y$. 

In this direction, let us consider a sequence where the jumping symbols are related to $\mu_j$
value ($\mu_j>3$) and the zero symbol is related other $\mu_z$ value. In this manner,
the flights have a characteristic scale while the rest periods may or may not have this characteristic scale
(depending on the $\mu_z$ value). The results concerning this scenario are shown in Figure ~\ref{fig:var2}b
where we show the dependence of $\alpha$ on $\mu_z$ for a fixed value of $\mu_j=6$ (different
values of $\mu_j>3$ does not affect this relation). From this figure,
we can identify three diffusive regimes: no diffusion for $\mu_z\approx1$ where the sequence is practically 
filled by zeros, subdiffusion $1\lesssim\mu_z\lesssim2$ and usual diffusion for $\mu_z\gtrsim 2 $.
}
\begin{figure}[!t]
\centering
\includegraphics[scale=0.31]{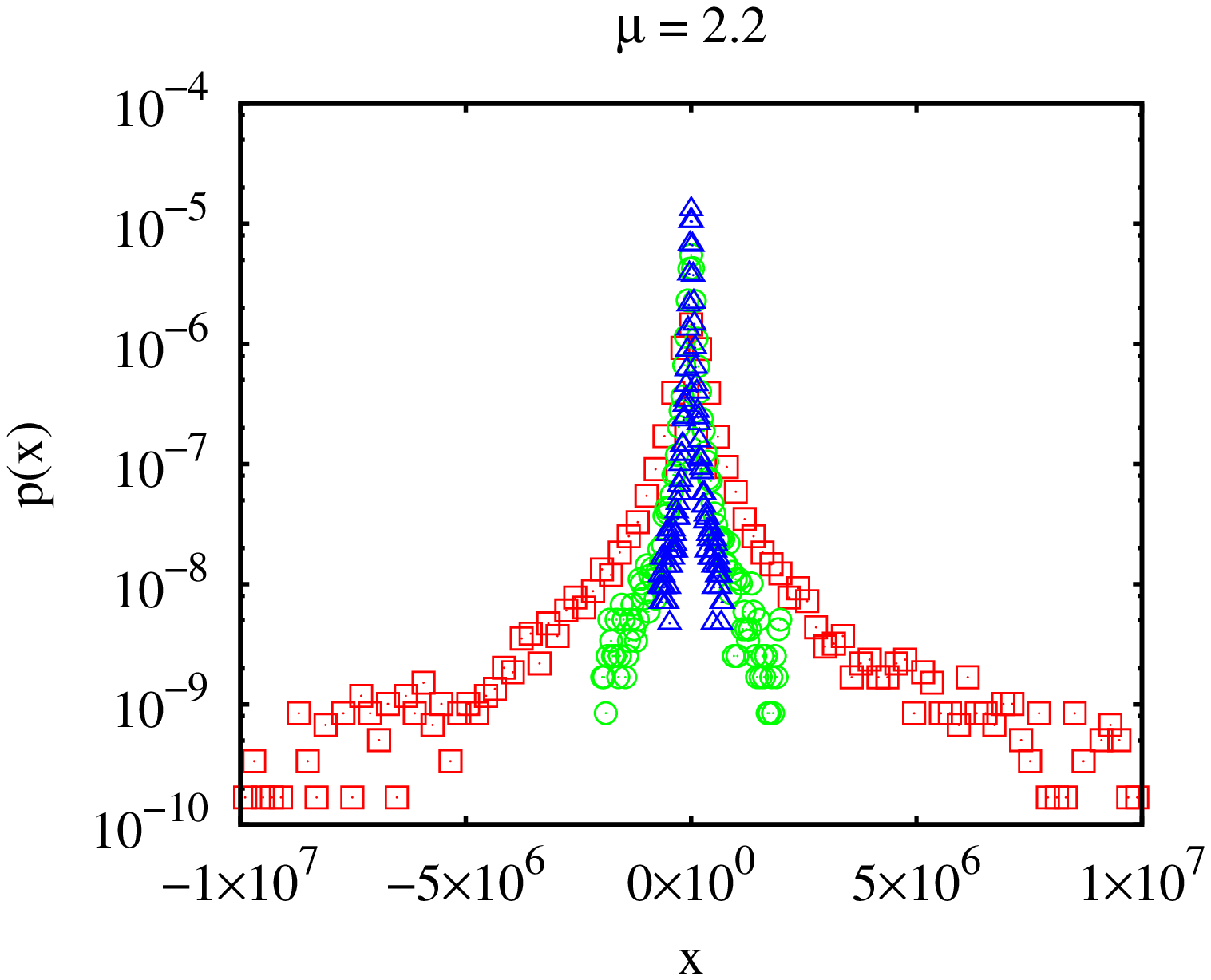}
\includegraphics[scale=0.31]{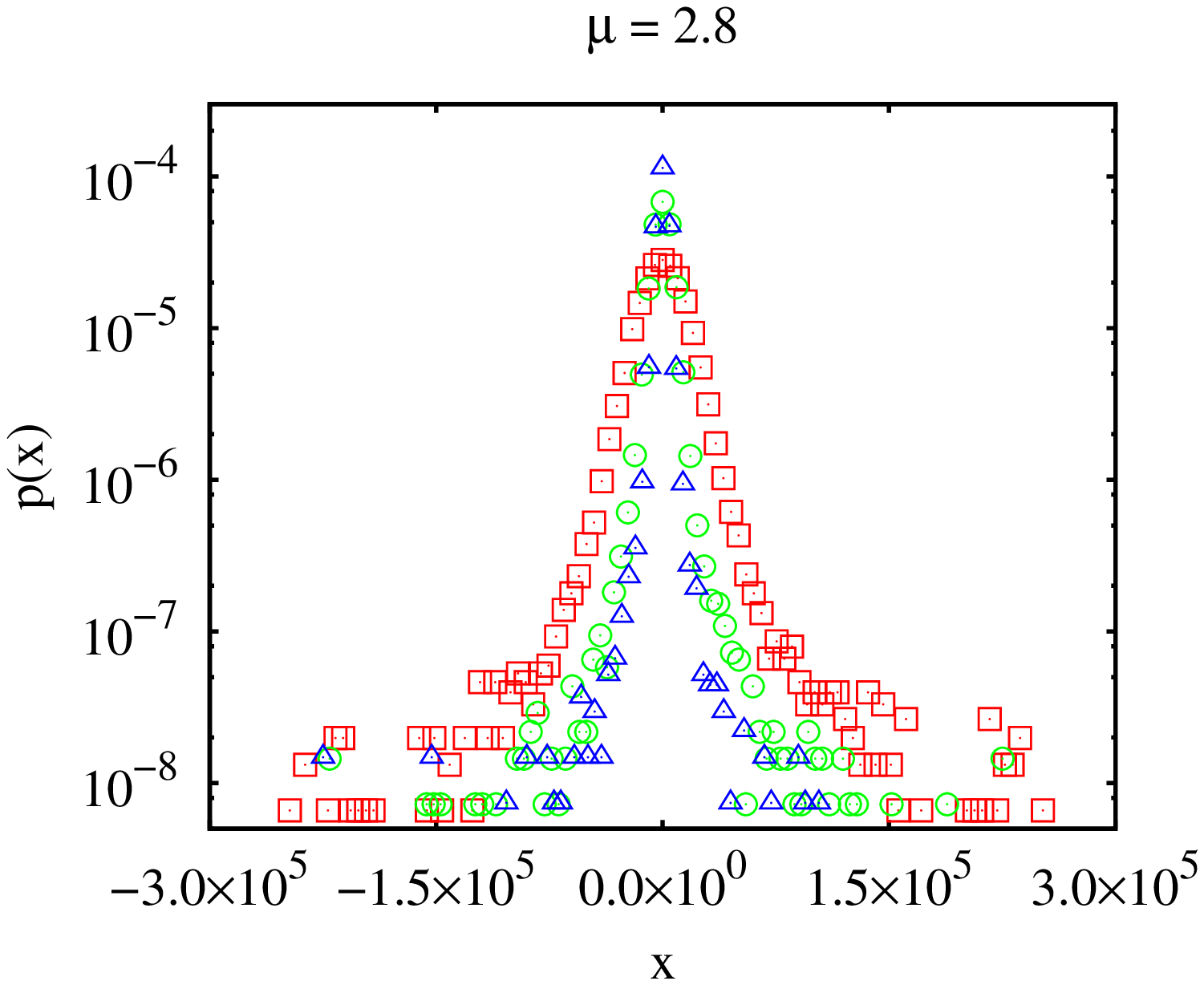}
\includegraphics[scale=0.31]{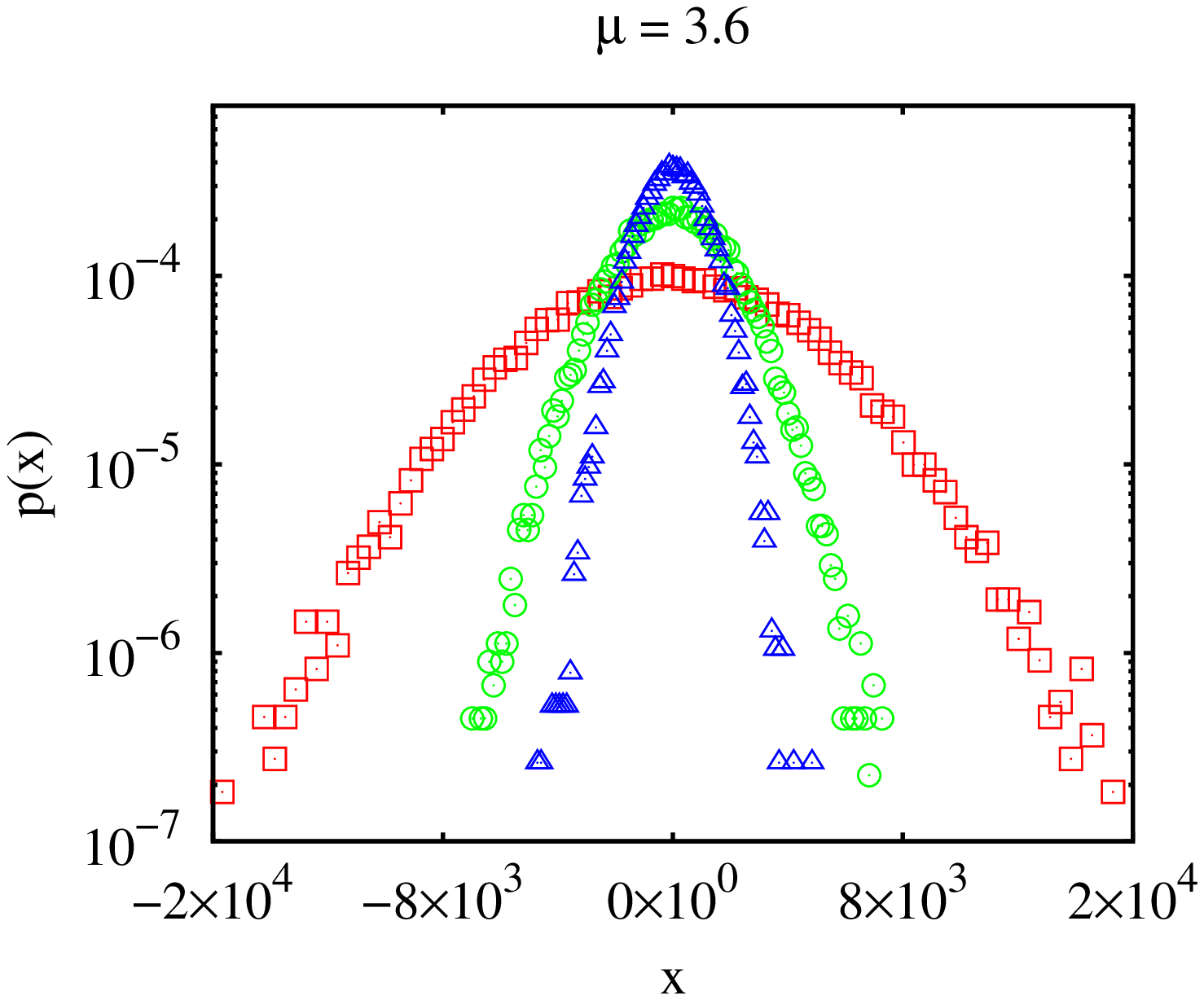}
\includegraphics[scale=0.31]{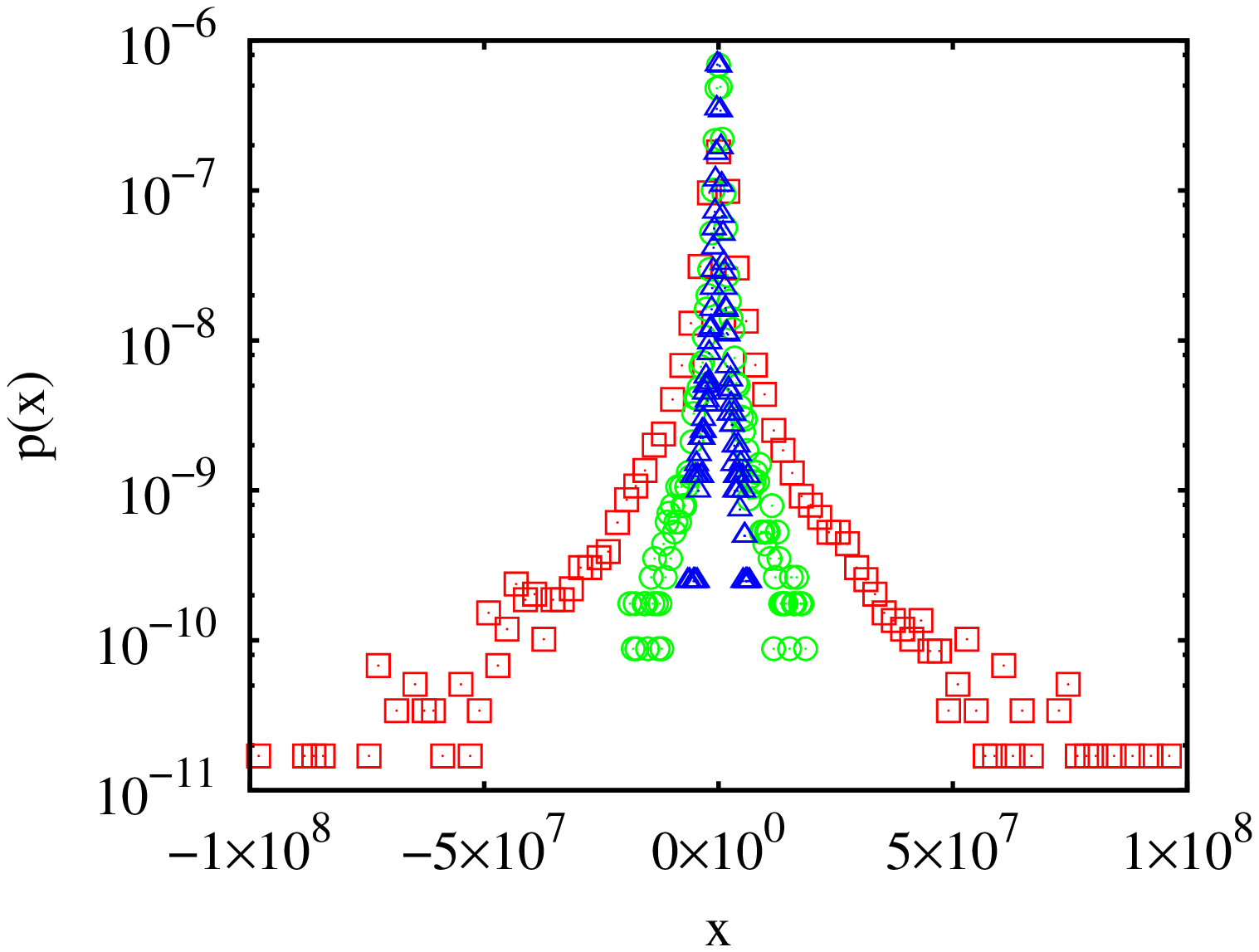}
\includegraphics[scale=0.31]{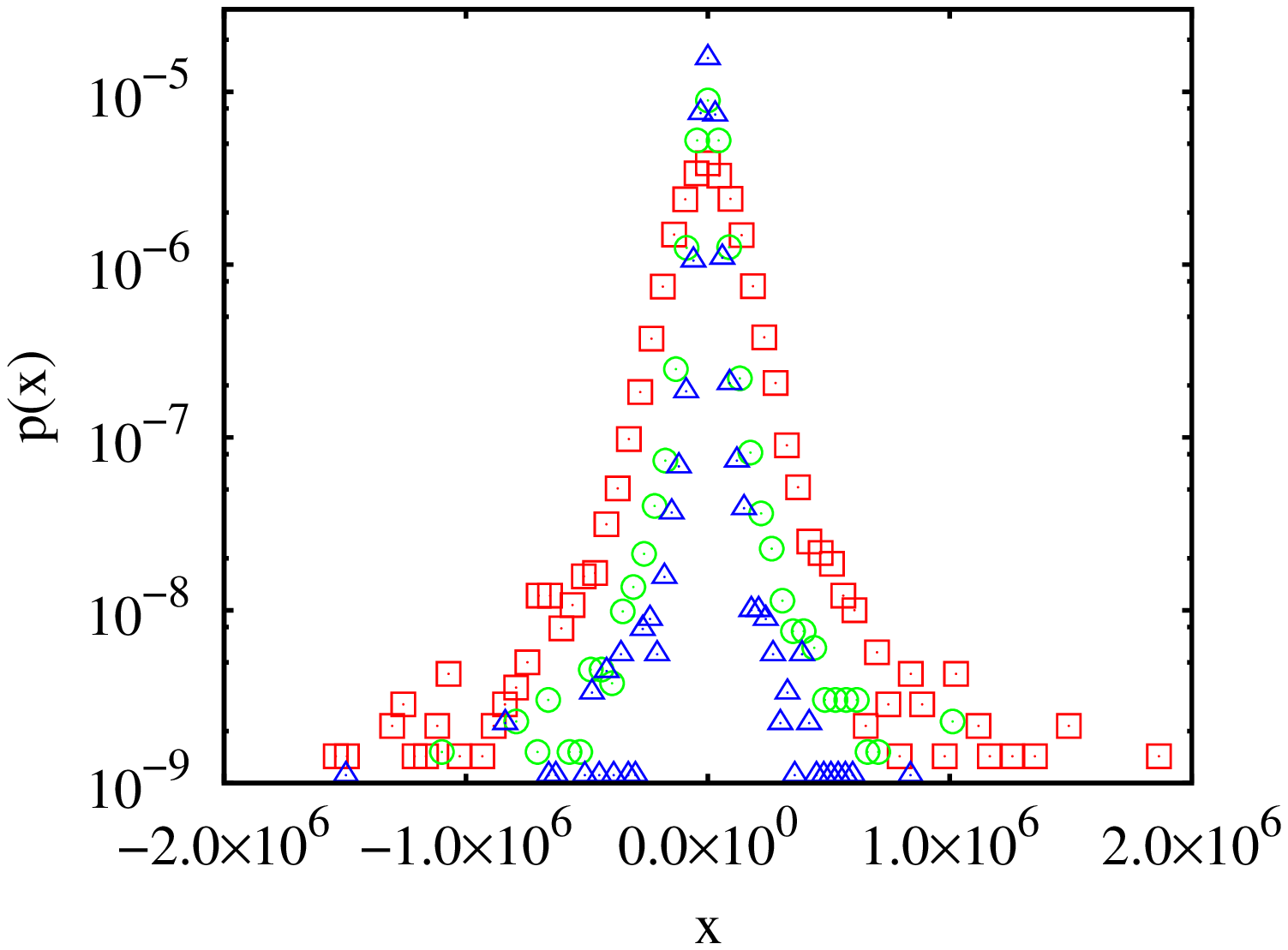}
\includegraphics[scale=0.31]{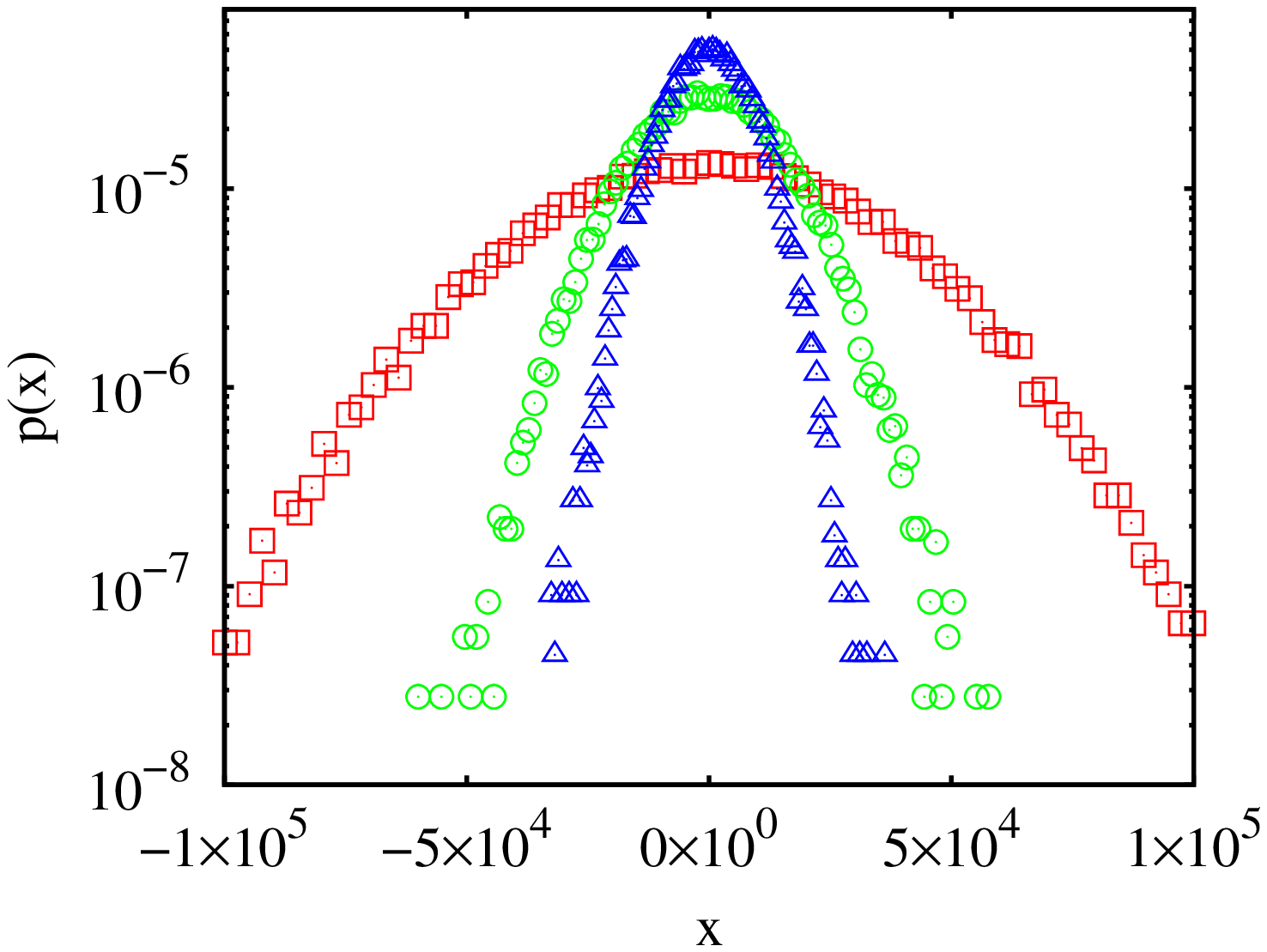}
\includegraphics[scale=0.31]{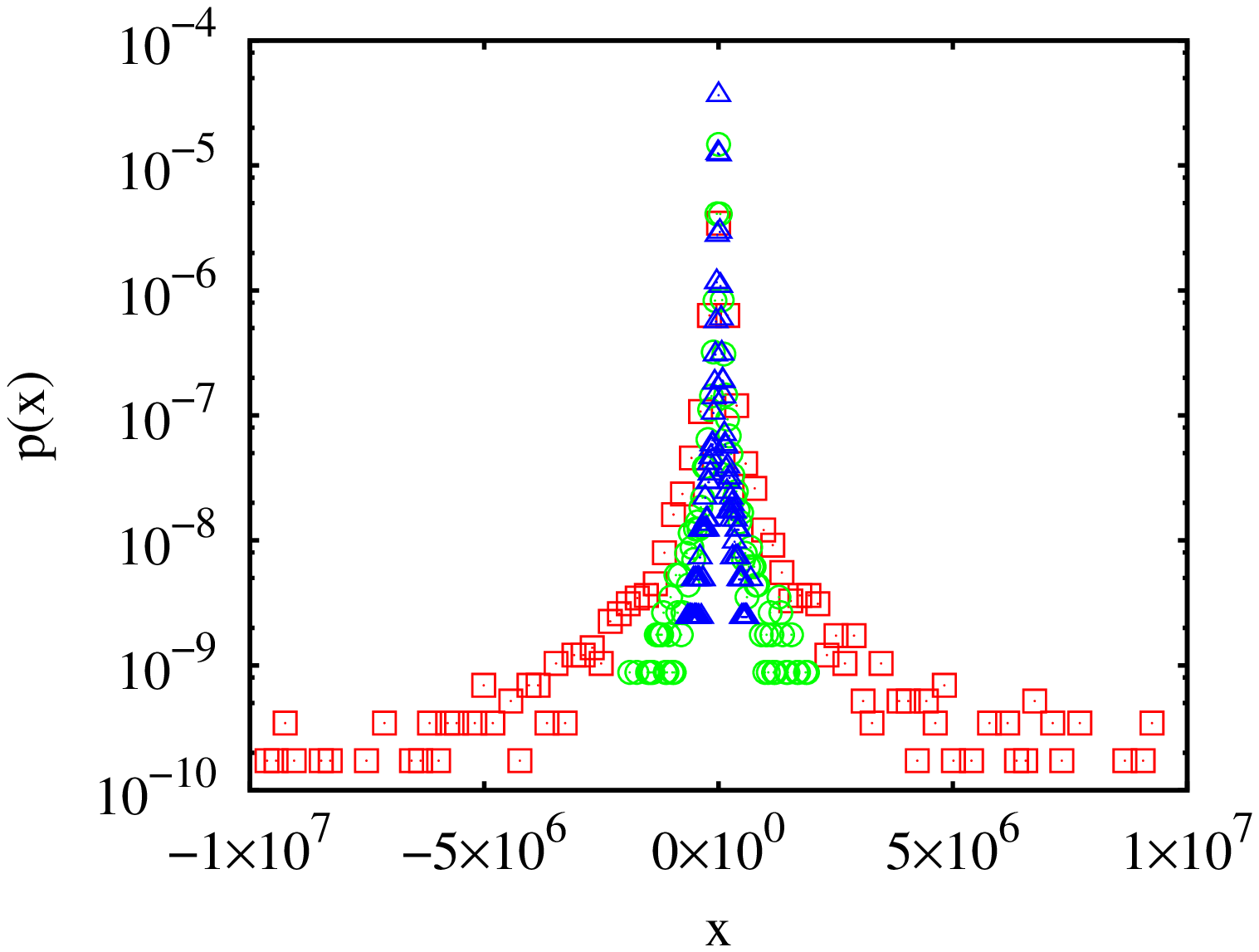}
\includegraphics[scale=0.31]{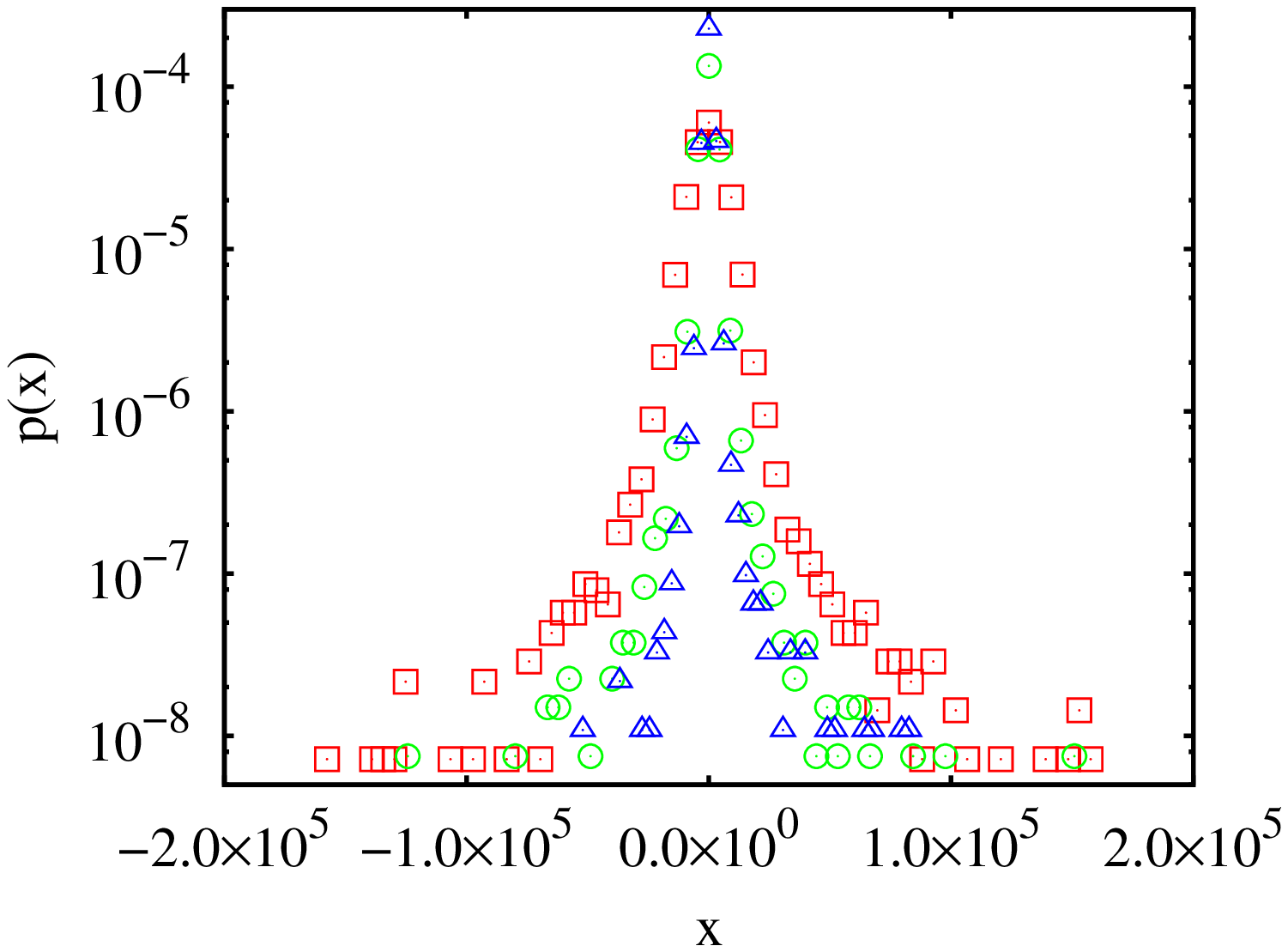}
\includegraphics[scale=0.31]{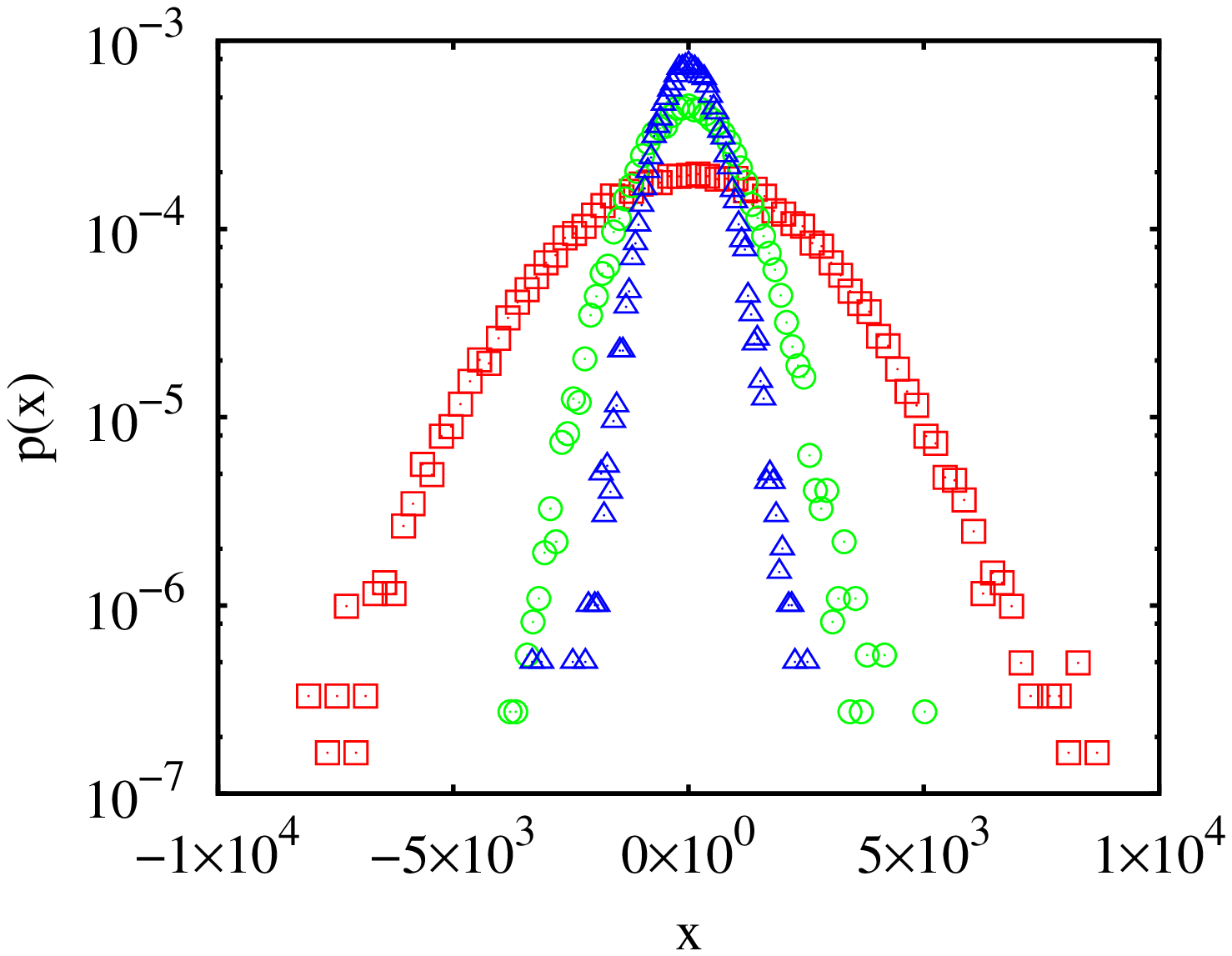}
\caption{Probability density function of $x(n)$ for the values of $\mu$ 
(indicated in the figures) for three values of $n$  $10^7$ (squares), 
$2\times 10^6$ (circles) and $714\times 10^3$ (triangles),
when considering the equiprobable alphabets  $\mathcal A=\{-1,1\}$ 
(upper panel), $\mathcal A=\{-11,\dots,0,\dots11\}$ (middle panel) and the 
alphabet $\mathcal A=\{-1,0^{10},1\}$ where the zero symbol is ten times more
probable than the others (lower panel). The histograms were obtained by 
using sequences of length $10^7$ with $A=1$ and $5\times10^5$
realizations of the numerical experiment.
}
\label{fig:hist}
\end{figure}
\begin{figure}[!ht]
\centering
\includegraphics[scale=0.31]{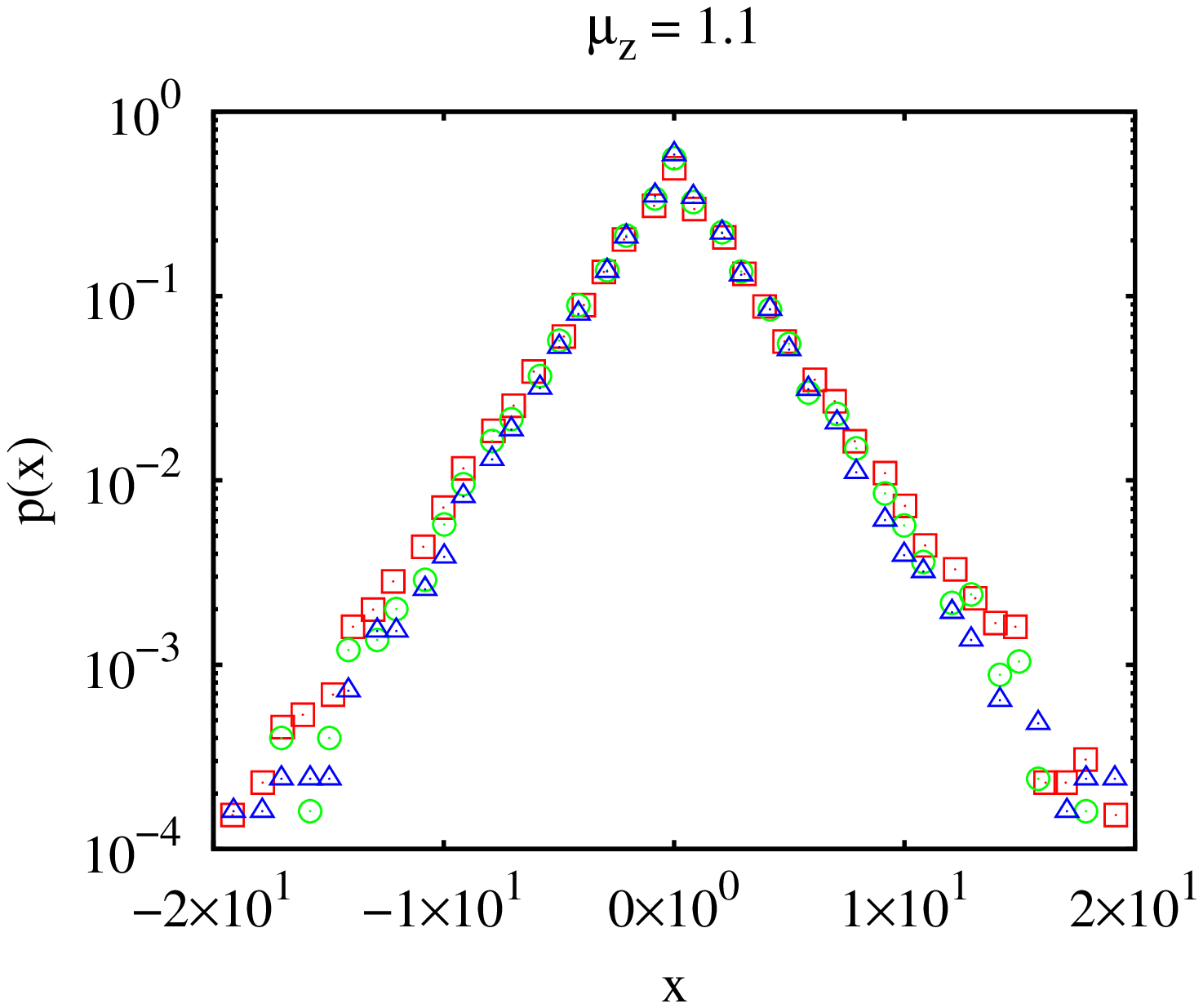}
\includegraphics[scale=0.31]{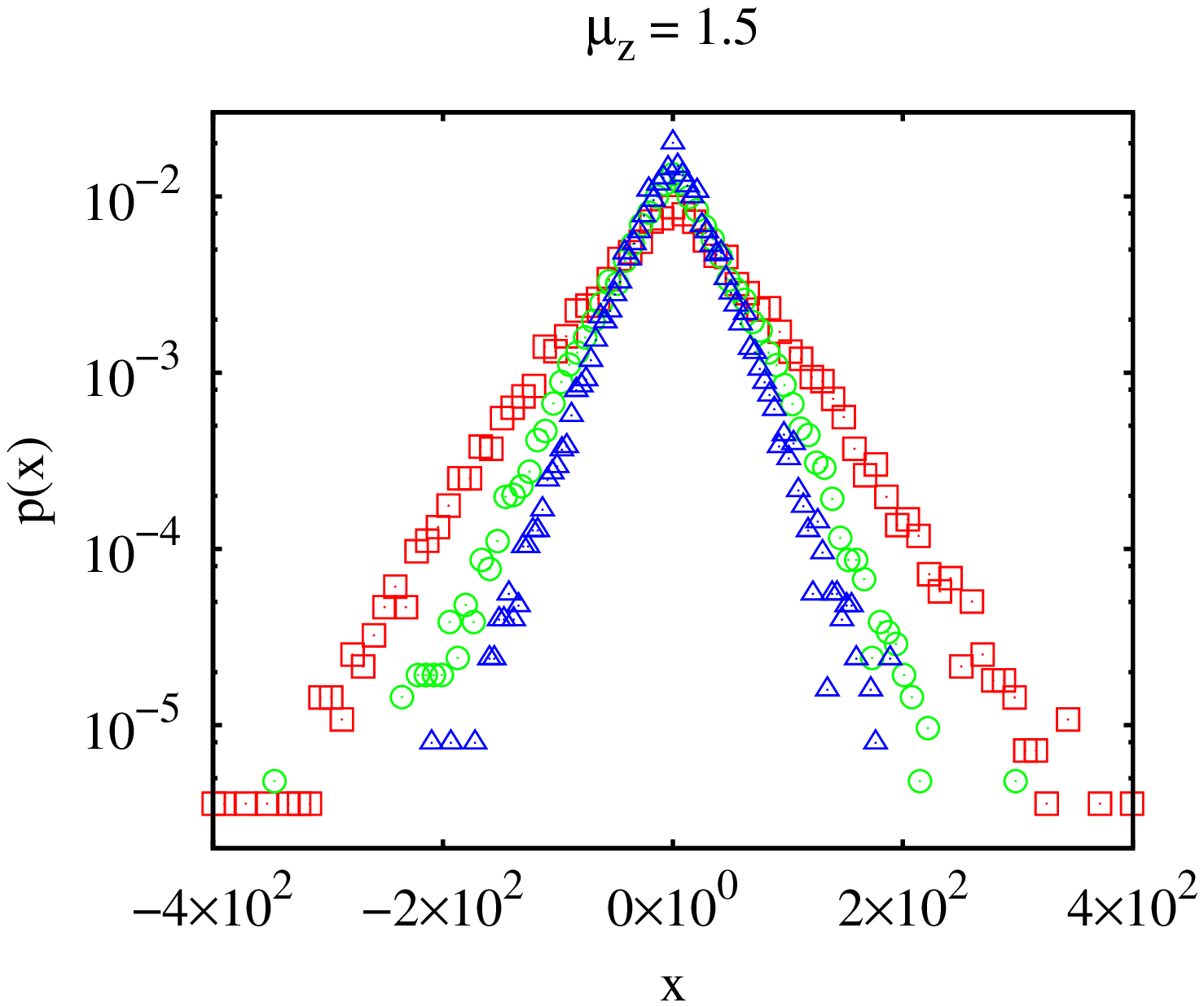}
\includegraphics[scale=0.31]{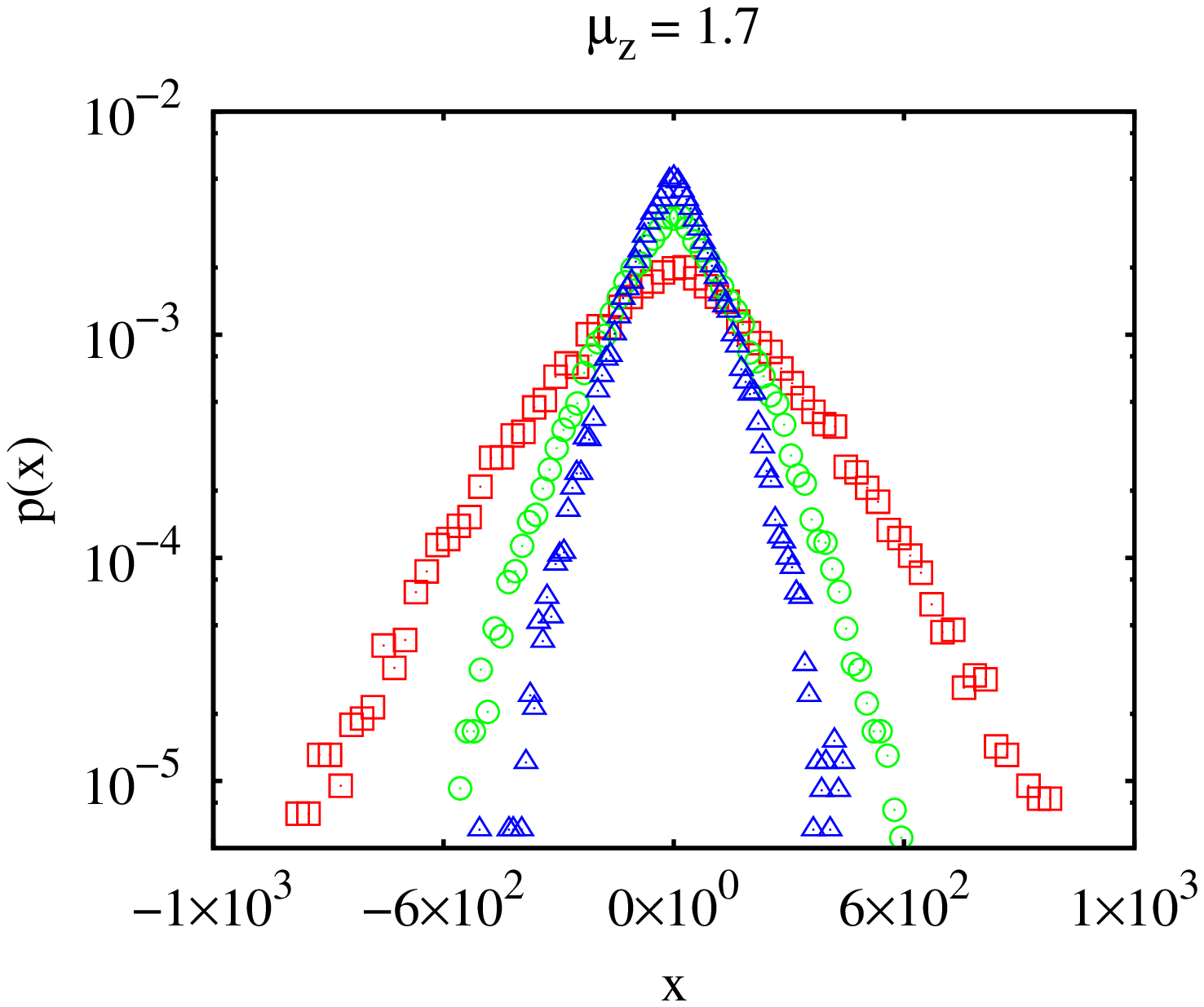}
\includegraphics[scale=0.31]{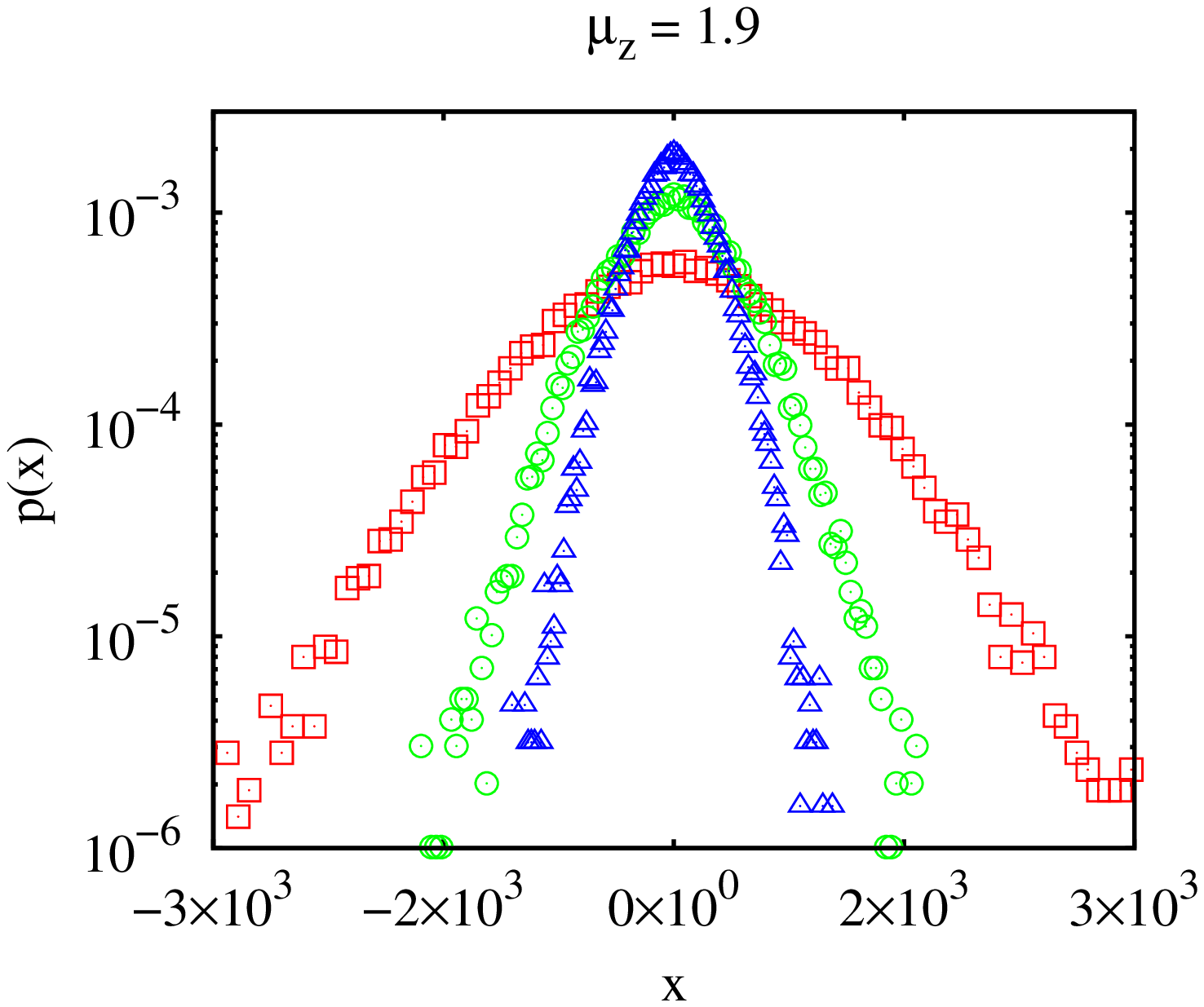}
\includegraphics[scale=0.31]{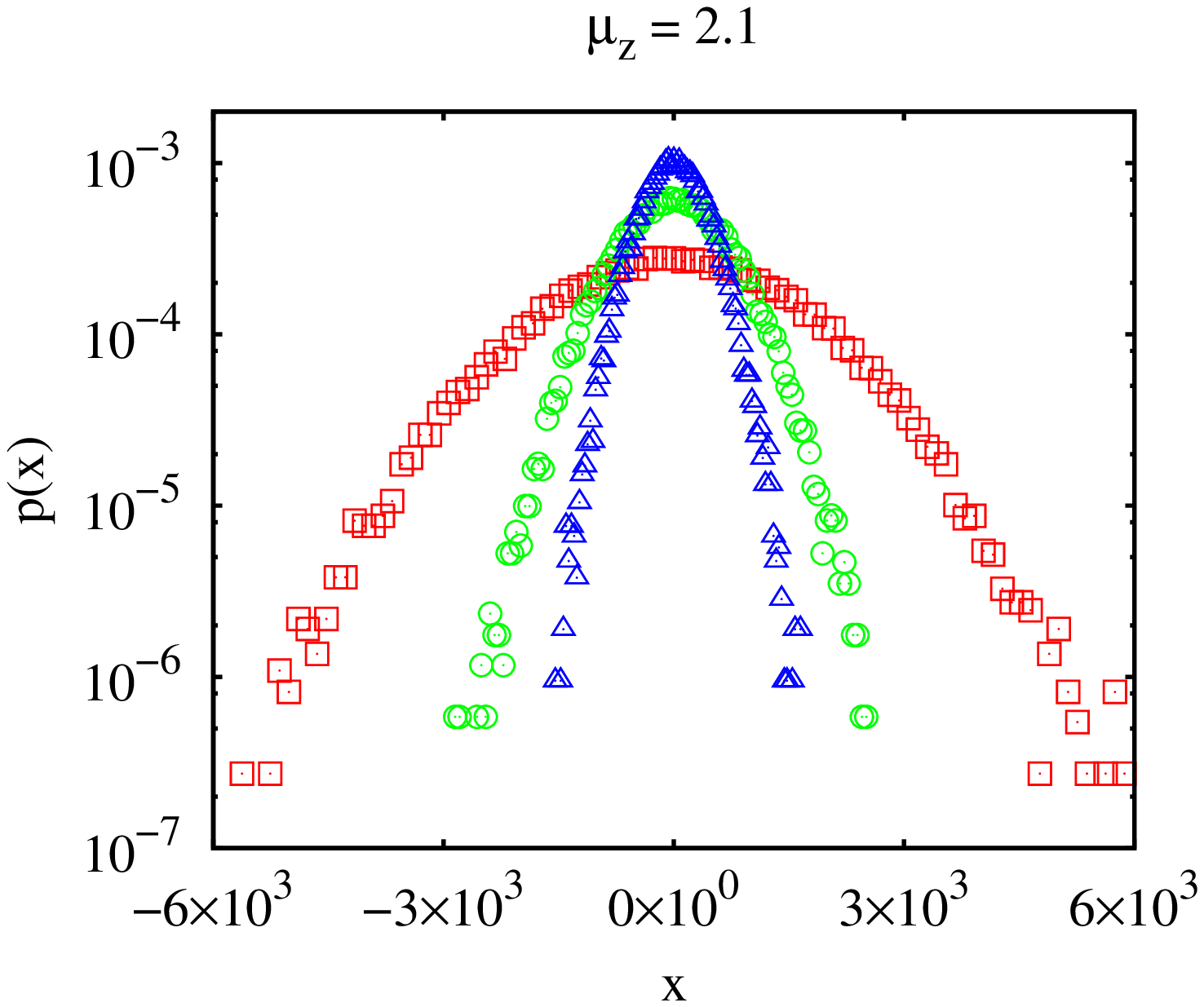}
\includegraphics[scale=0.31]{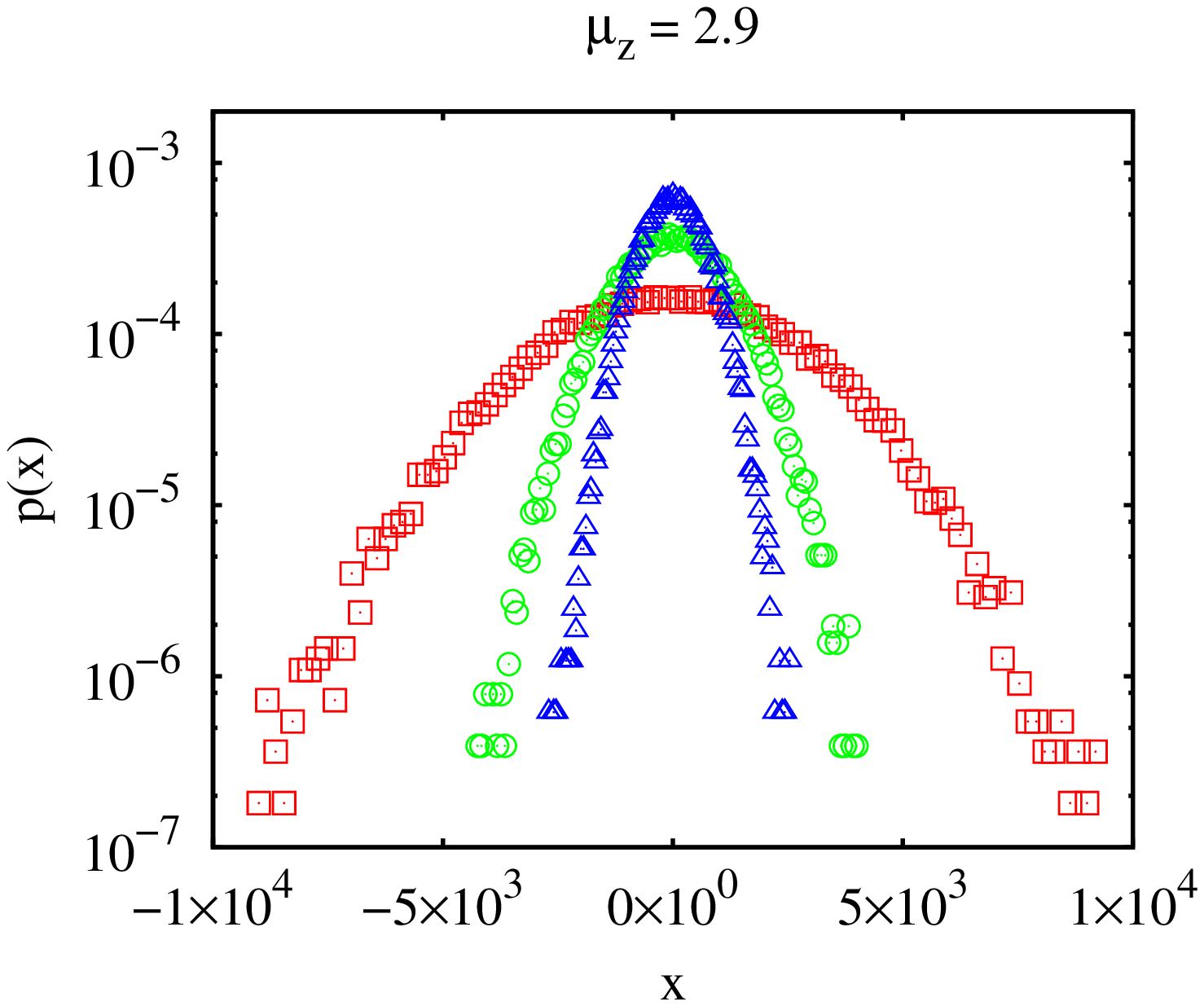}
\caption{Probability density function of $x(n)$ for the values of $\mu_z$ 
(indicated in the figures) for three values of $n$  $10^7$ (squares), 
$2\times 10^6$ (circles) and $714\times 10^3$ (triangles).
The histograms were obtained by using sequences of length $10^7$ 
with $A=1$, $\mu_j=6$ and $5\times10^5$ realizations of the numerical experiment
with the alphabet $\mathcal A=\{-1,0,1\}$.}
\label{fig:histz}
\end{figure}

We also evaluated the probability density functions (pdf) of $x(n)$  to investigate 
the shape of distribution $p(x,n)$ for different values of $\mu$ {as well as
for the different constructions of the erratic trajectories. Figure \ref{fig:hist}
shows these distributions for the equiprobable alphabets  $\mathcal A=\{-1,1\}$ 
(upper panel), $\mathcal A=\{-11,\dots,0,\dots11\}$ (middle panel) and the 
alphabet $\mathcal A=\{-1,0^{10},1\}$ with the zero symbol being ten times more
probable than the others (lower panel).} We can see that the distributions are characterized 
by non-Gaussian profiles with heavy tails when $\mu\lesssim3$, recovering
the Gaussian propagator when $\mu\gtrsim 3$. {Further, a visual inspection
suggests that the different constructions of the erratic trajectories only change the
scale of these plots.

The situation is remarkably different when considering one value of $\mu$ for
the jumping symbols and other for the zero symbol with the alphabet $\mathcal A=\{-1,0,1\}$. 
Figure \ref{fig:histz} shows the distributions for this case. Notice that the shape of distributions goes from a 
Laplace ($p(x)\sim \exp(-|x|)$) to Gaussian distribution, depending on the $\mu_z$ value.
}

\section{Continuous-time random walk models}

{
So far we have empirically described the diffusive behavior of
the symbolic sequences proposed by Buiatti \textit{et al.}. Now
let us compare these empirical findings with some analytical 
models based on  continuous-time random walk. 
}

In the continuous-time random walk (CTRW) of Montroll\cite{Montroll} (see also \cite{Metzler}), the random walk 
process is fully specified by the function $\psi(x,t)$, the probability density  to 
move a distance $x$ in  time $t$. We can distinguish three different ways to make 
the movement: the particle waits until it moves instantaneously to a new position 
(jump model) or the particle moves at constant velocity to a new position and chooses 
randomly a new direction (velocity model) or the particle moves at constant velocity 
between turning points that are chosen randomly\cite{Zumofen}. {There are two
fundamental approaches to the CTRW: (i) the decoupled and (ii) the coupled 
formalisms. In (i) the function $\psi(x,t)$ is supposed to factor in the form 
$\psi(x,t)=w(t) \lambda(x)$, i.e., the jumping and the waiting time are independent
random variables. For (ii) both process are coupled, a jump of a certain length may
involve a time cost or vice-versa. This coupled form commonly leads to more
cumbersome calculations. 

Here, we notice that because of the erratic trajectories construction, every
continuous jump (without changing the sequence symbol) with length $N_y$
occurs at constant velocity and costs the same $N_y$ units of time to be performed. 
This fact lead us to the velocity and coupled model when considering
the equiprobable alphabet $\mathcal A=\{-1,1\}$. When adding the zero symbol
the resting times are decoupled from the jumps, but the jumping times stay coupled.
In addition, we have to remember that our formal time $n$ is a discrete variable. 
Thus, comparisons with this formalism should be viewed as semi-quantitative.
In this context, it is interesting to note that the work of Gorenflo \textit{et al.}\cite{Gorenflo, Gorenflo2} 
extends the Montroll's theory to the discrete domain considering the decoupled version of the CTRW,
in contrast to the first approach used here.
}

\begin{figure}[!ht]
\centering
\includegraphics[scale=0.31]{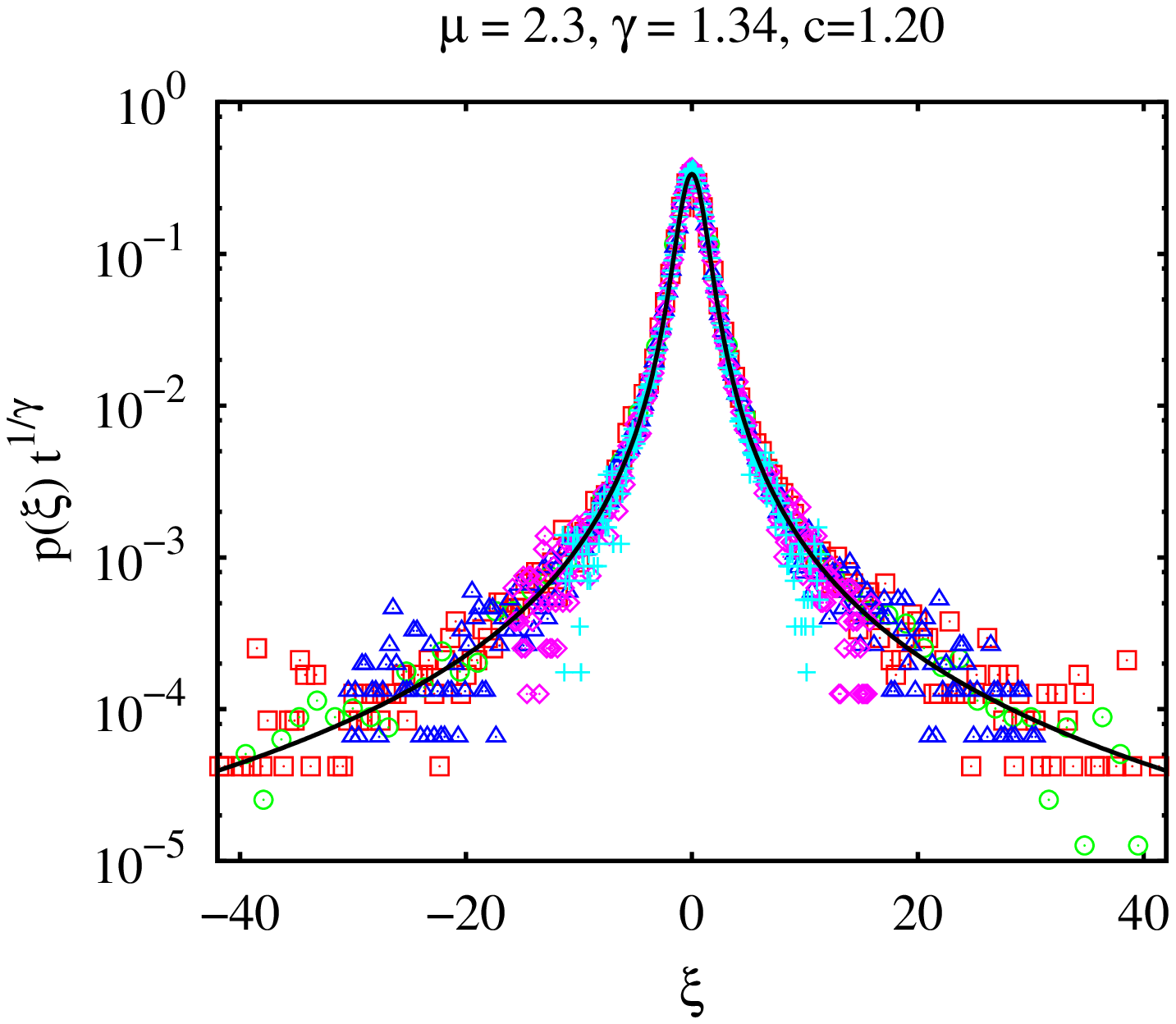}
\includegraphics[scale=0.31]{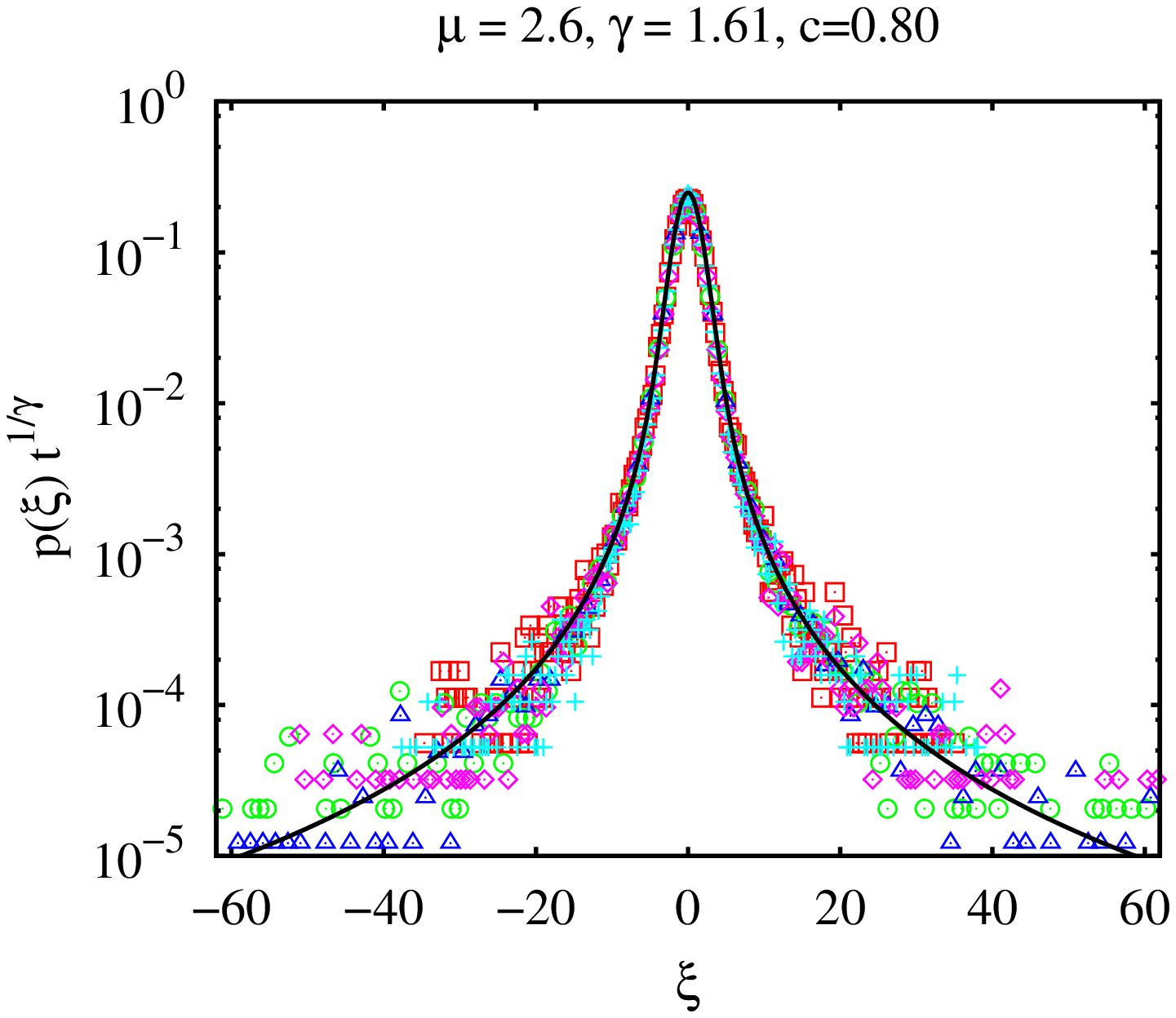}
\includegraphics[scale=0.31]{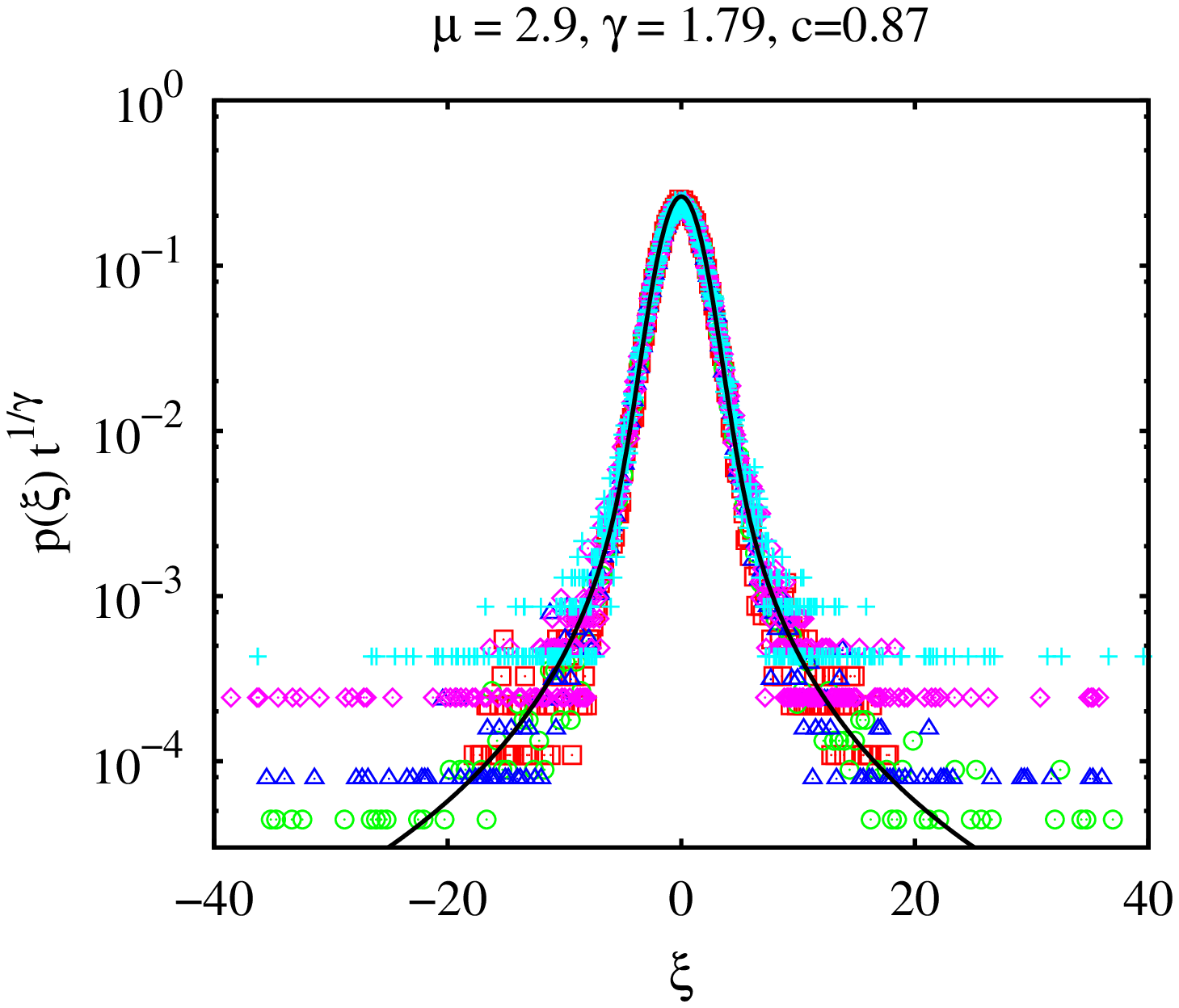}
\includegraphics[scale=0.31]{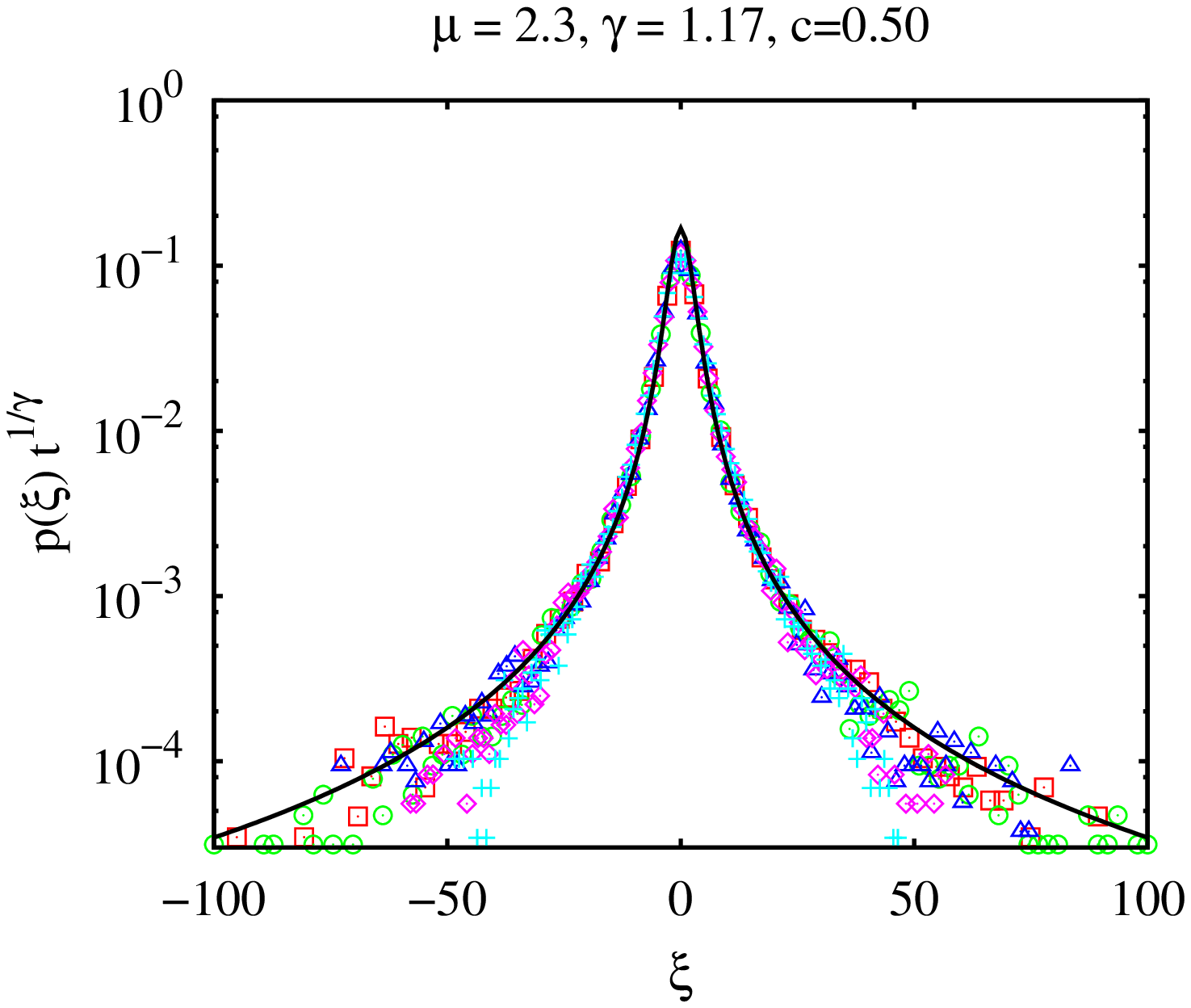}
\includegraphics[scale=0.31]{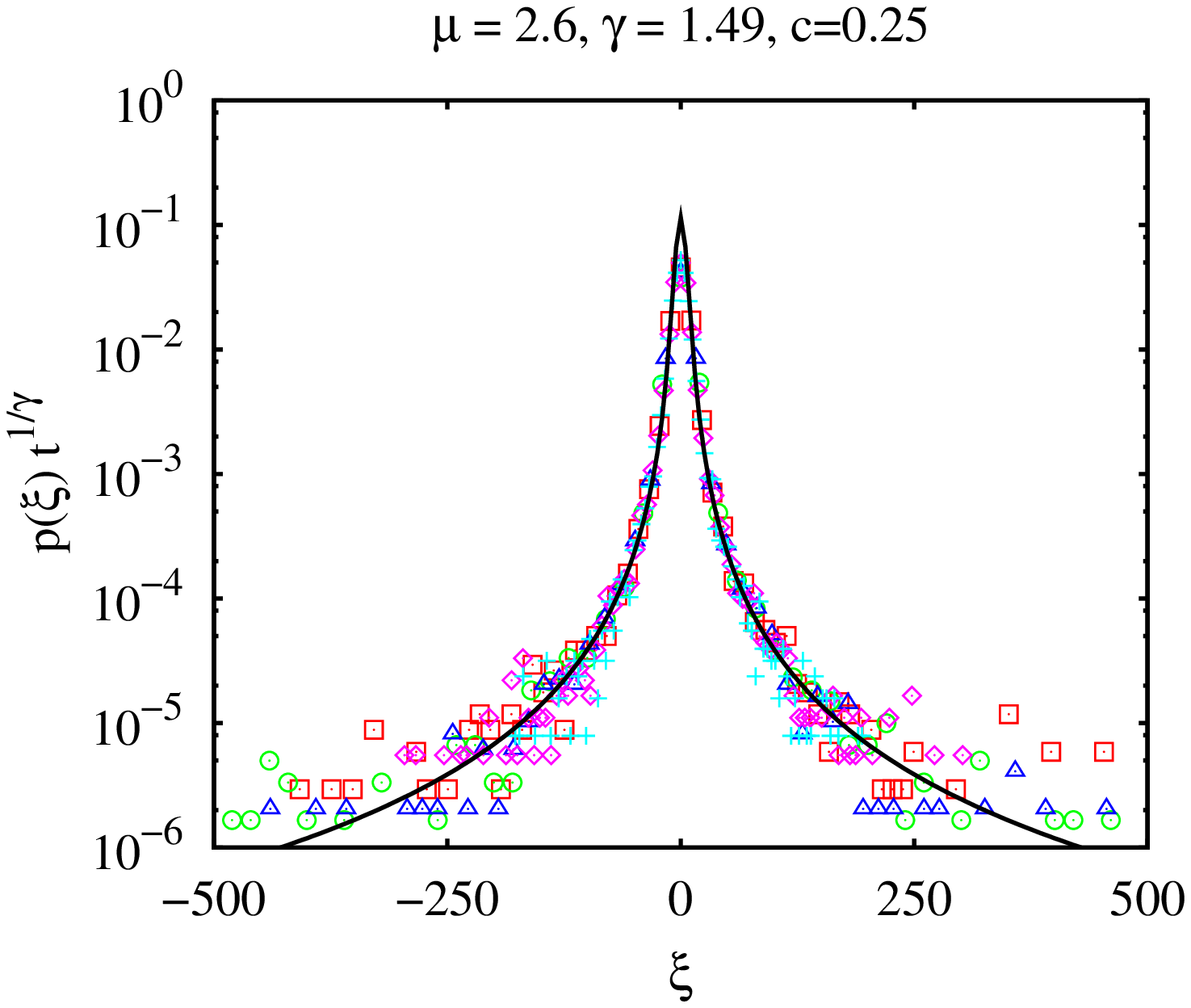}
\includegraphics[scale=0.31]{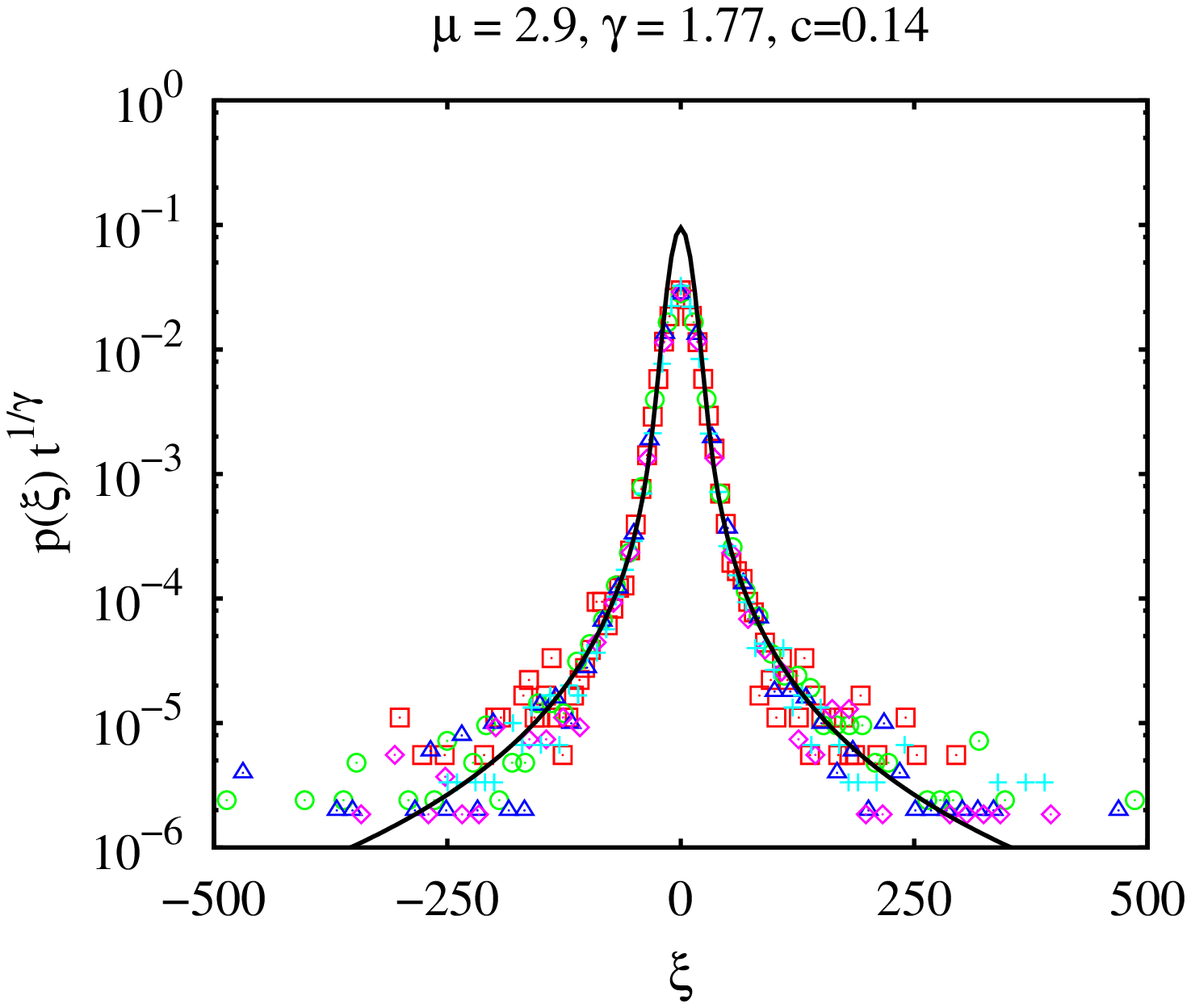}
\includegraphics[scale=0.31]{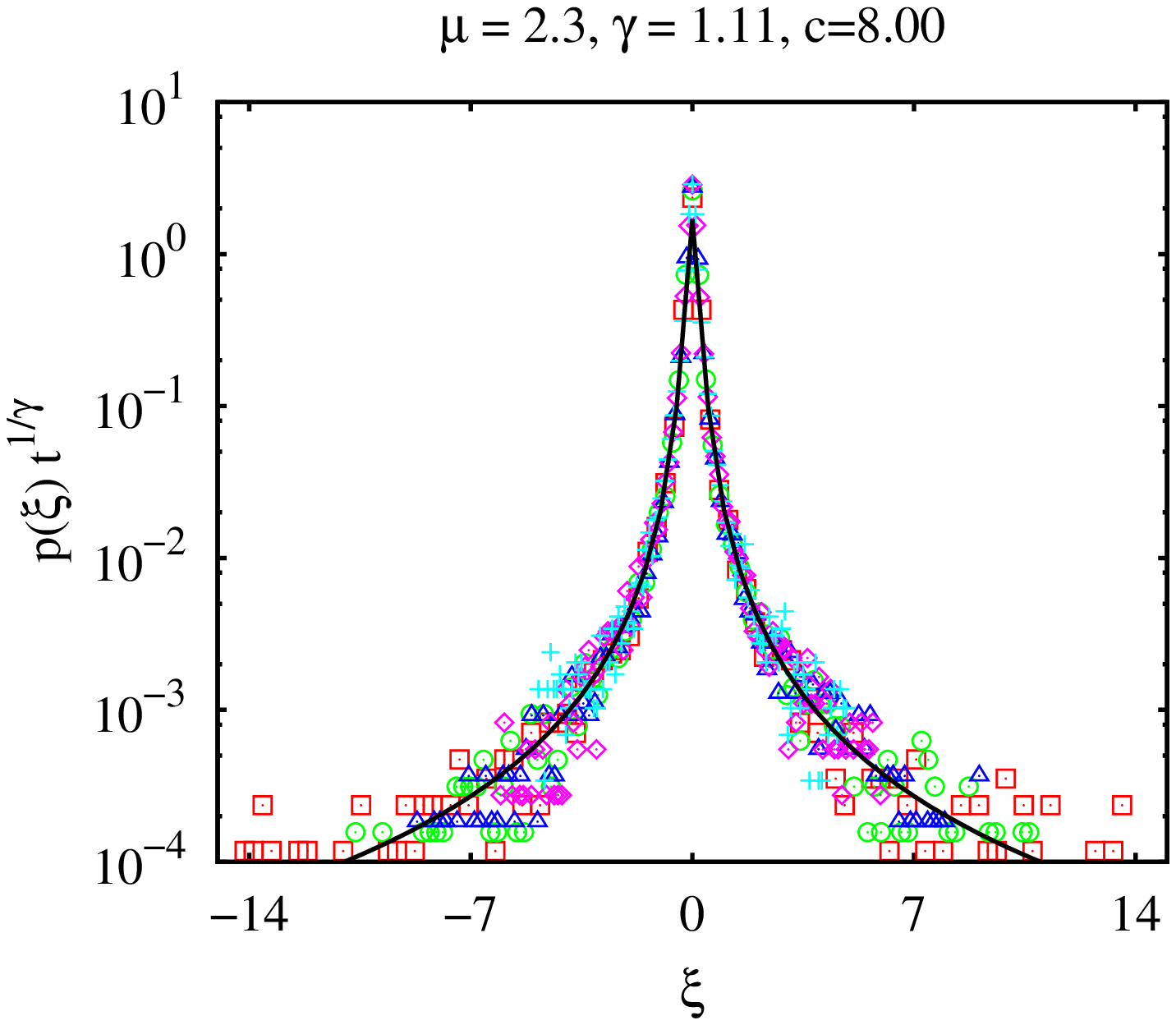}
\includegraphics[scale=0.31]{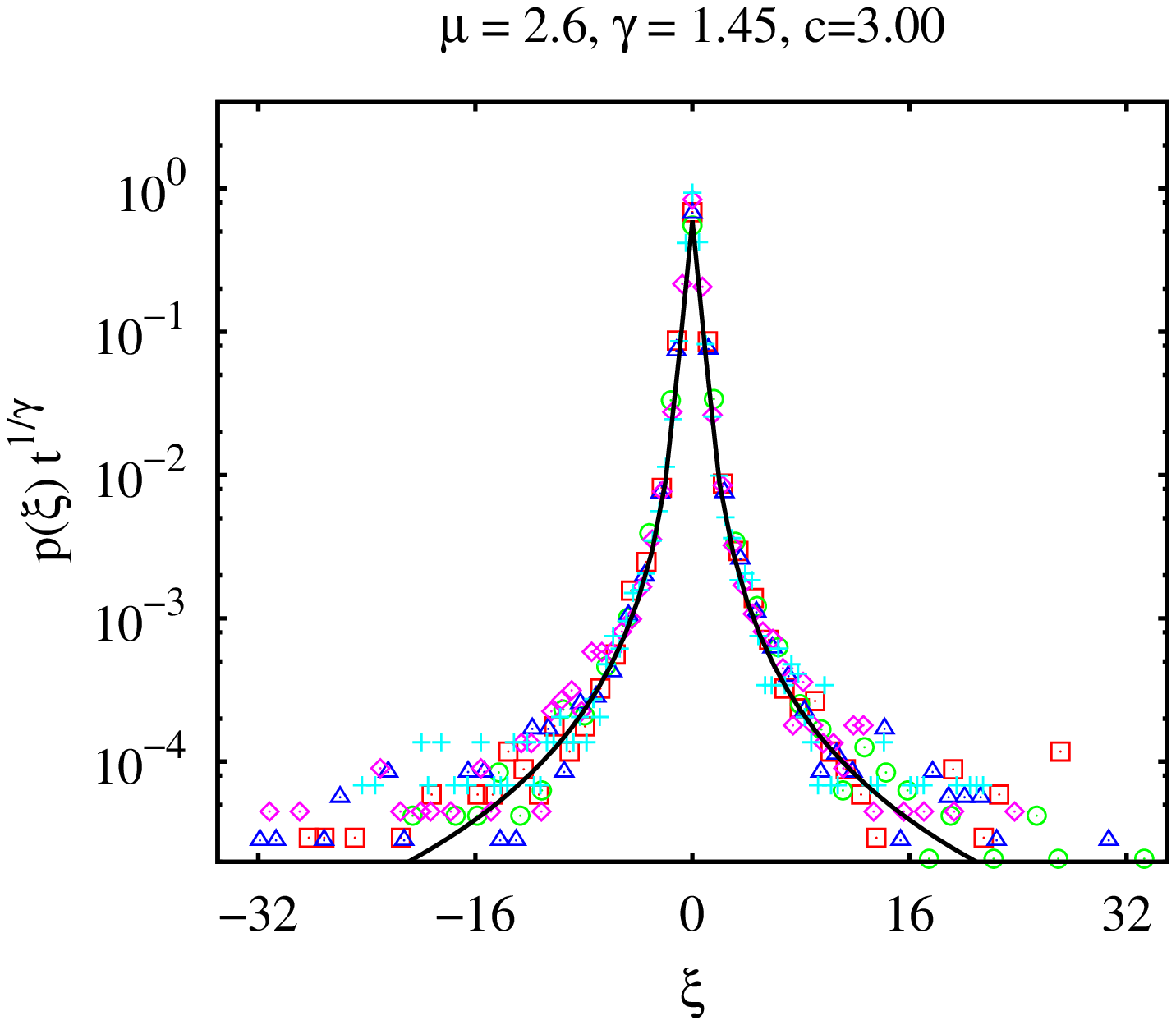}
\includegraphics[scale=0.31]{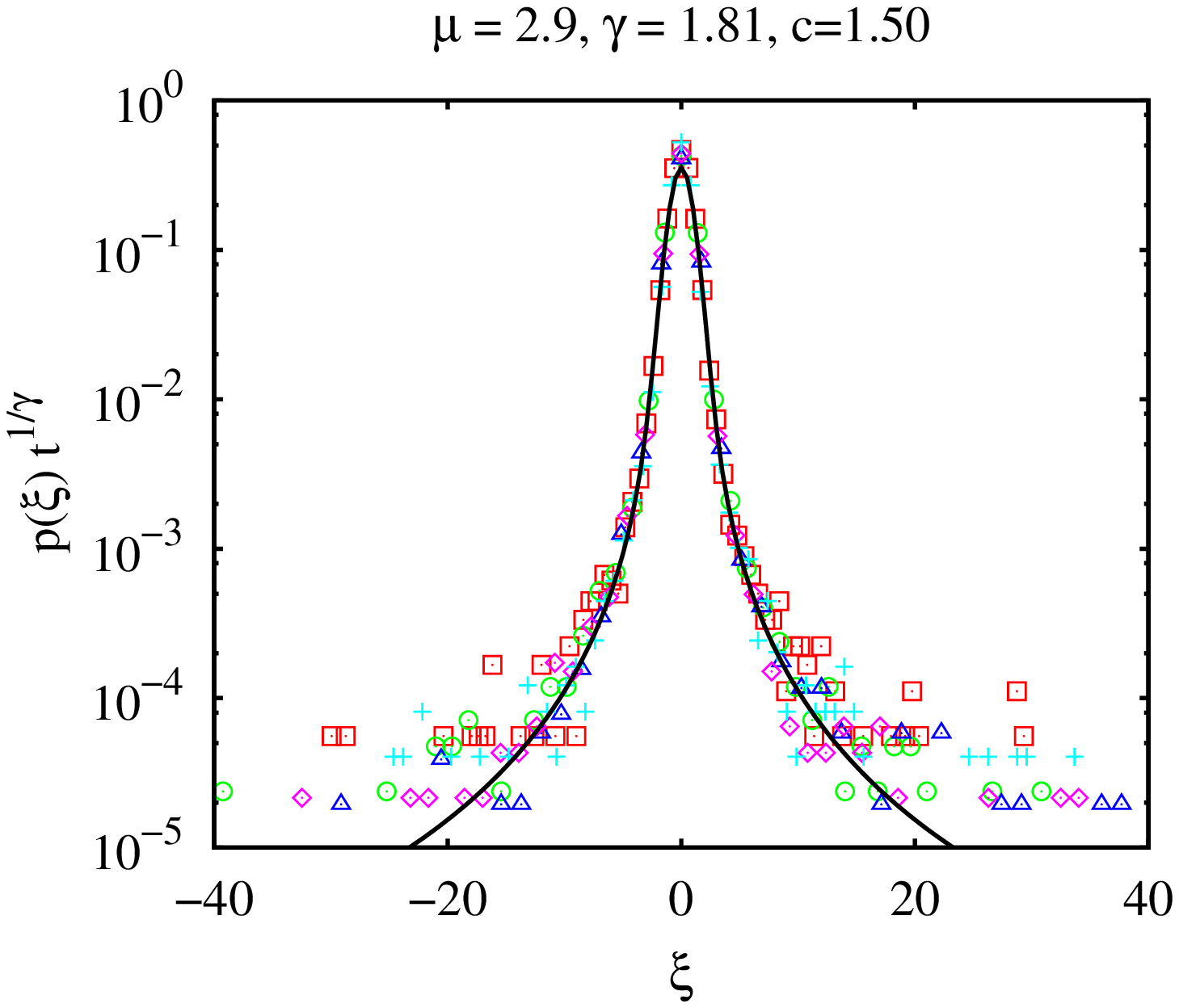}
\caption{Probability density function of the scaling variable $\xi=c\, x/t^\gamma$ for some 
values of $\mu$ (indicated in the figure) for five values of $n$:  $10^7$ (squares), 
$2\times 10^6$ (circles), $714\times 10^3$ (triangles), $55\times 10^3$ (diamonds)
and $15\times 10^3$ (crosses). The upper panel shows the results for the 
equiprobable alphabet  $\mathcal A=\{-1,1\}$, the middle panel for 
$\mathcal A=\{-11,\dots,0,\dots11\}$ and the lower panel for $\mathcal A=\{-1,0^{10},1\}$ with 
the zero symbol been ten times more probable than the others.
The continuous lines are the predictions of the the continuous-time random walk model, 
equation (\ref{pdf}). The values of $c$ and $\gamma$ are indicated in the figure and
the numerical data was obtained from $5\times 10^{5}$ realizations 
of equation (\ref{eq:pe}) with $A=1$.}
\label{fig:histc}
\end{figure}

\begin{figure}[!ht]
\centering
\includegraphics[scale=0.31]{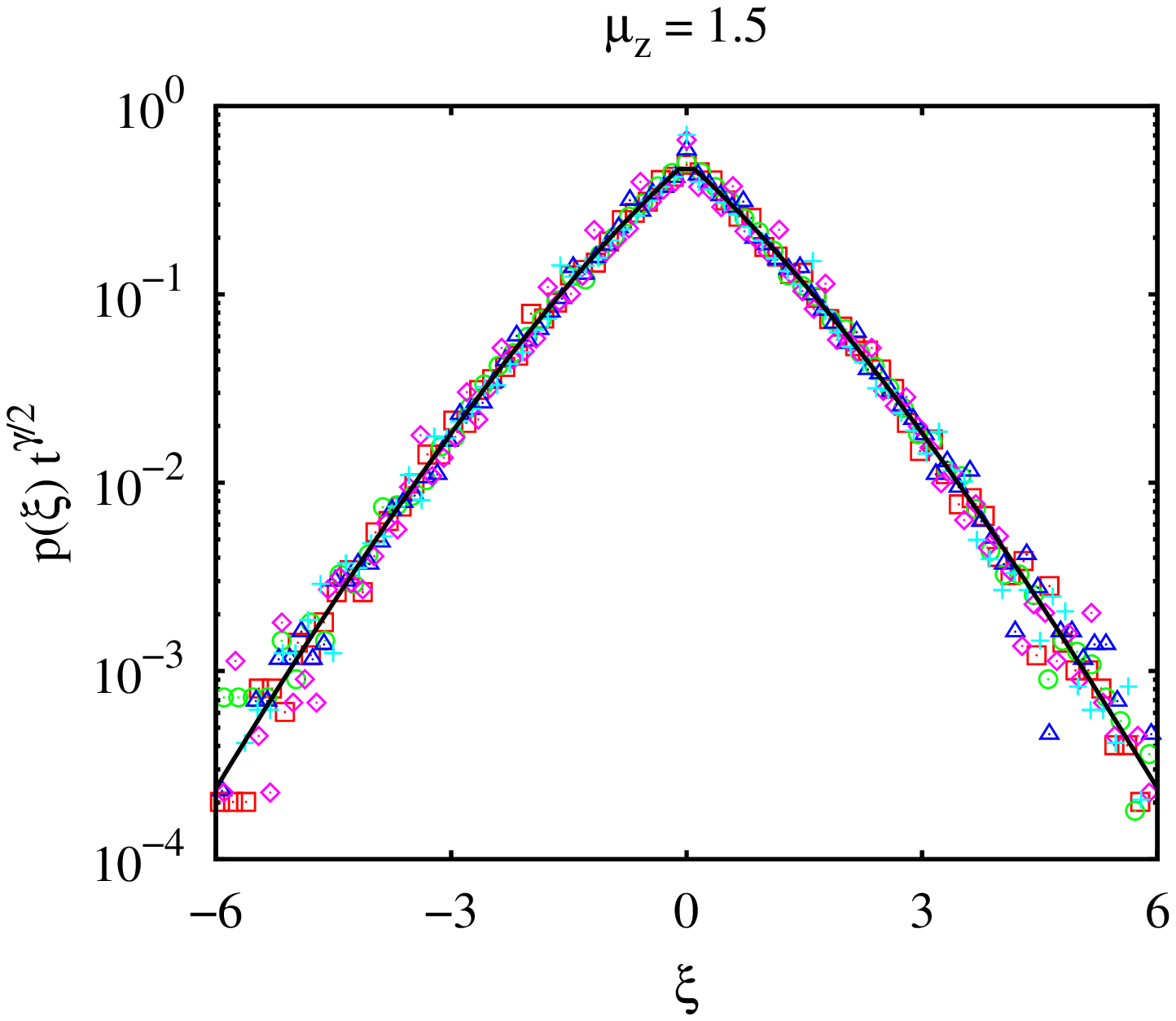}
\includegraphics[scale=0.31]{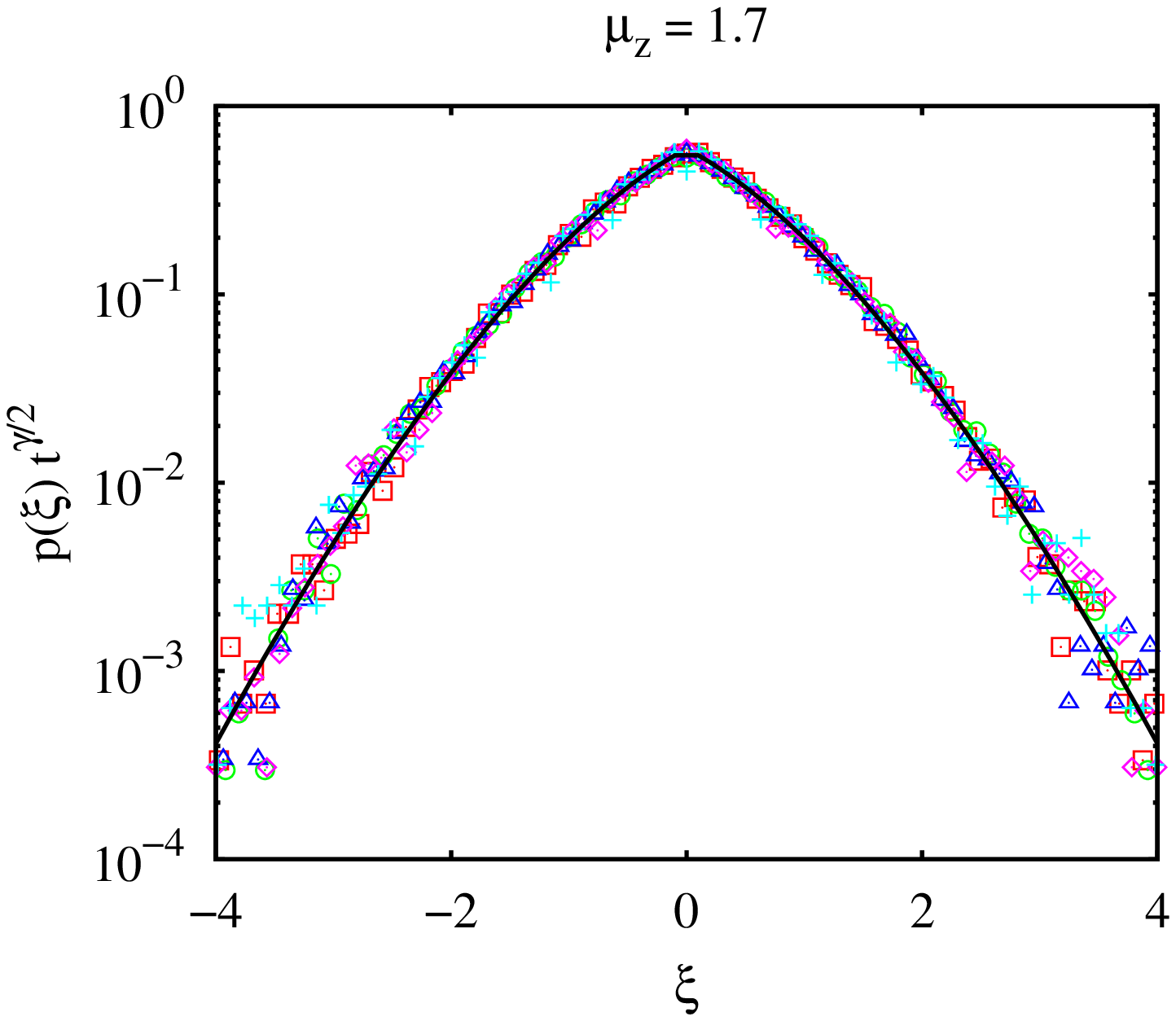}
\includegraphics[scale=0.31]{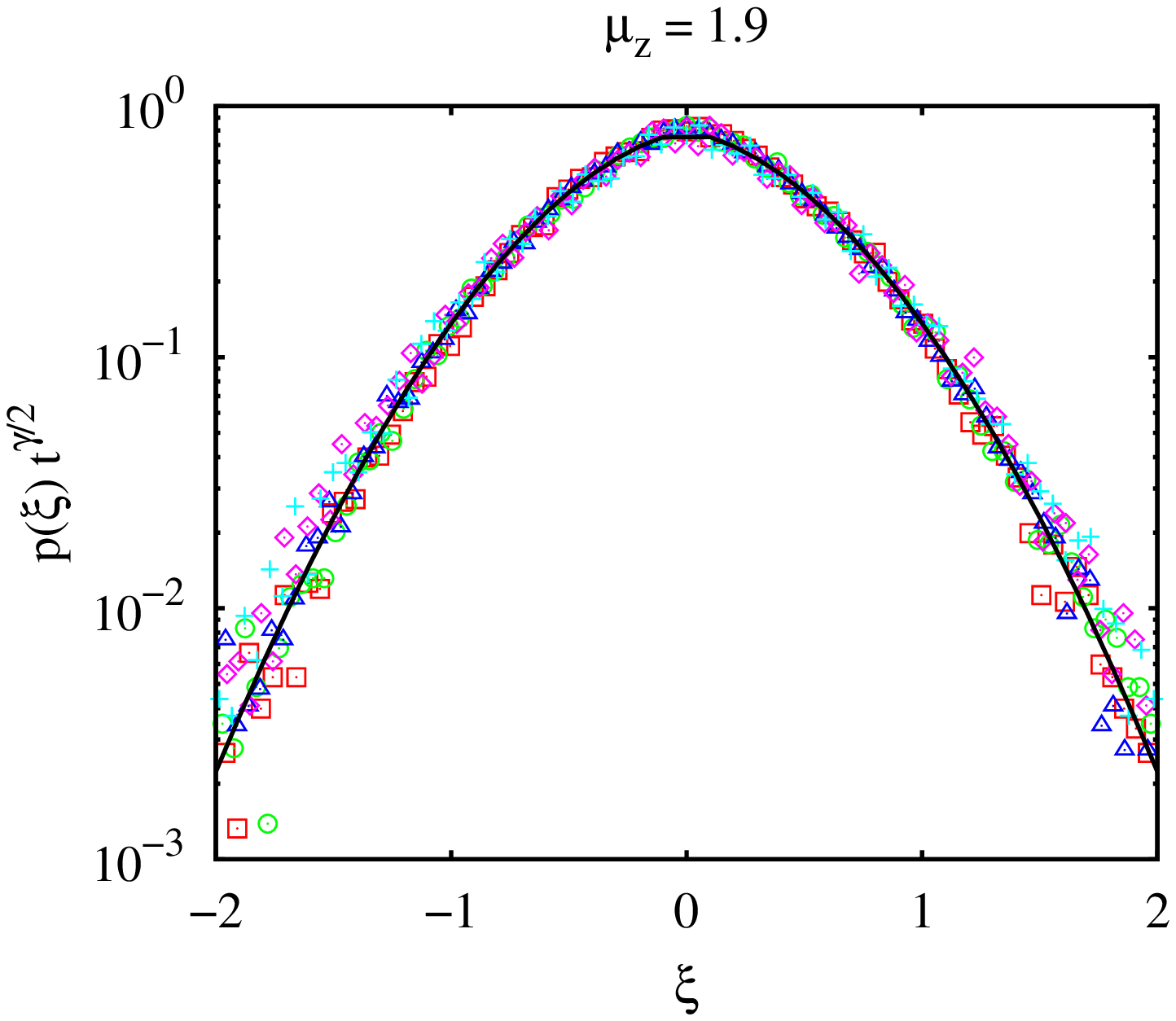}
\caption{Probability density function of the scaling variable $\xi=c x/t^{\gamma/2}$ for some 
values of $\mu_z$ (indicated in the figure) for five values of $n$:  $10^7$ (squares), 
$2\times 10^6$ (circles), $714\times 10^3$ (triangles), $55\times 10^3$ (diamonds)
and $15\times 10^3$ (crosses) when considering the equiprobable alphabet 
$\mathcal A=\{-1,0,1\}$ and $\mu_j=6$. The continuous line is the prediction of the the 
continuous-time random walk model, equation (\ref{pdf2}) with $\gamma=0.47 $ ($c=0.65$) 
for $\mu_z=1.5$,  $\gamma=0.64$ ($c=0.37$) for $\mu_z=1.7$,  $\gamma=0.94$ ($c=0.15$) 
for $\mu_z=1.9$. The numerical data was obtained from $5\times 10^{5}$ realizations of the numerical 
experiment with $A=1$.}
\label{fig:histcd}
\end{figure}

\begin{figure}[!ht]
\centering
\includegraphics[scale=0.4]{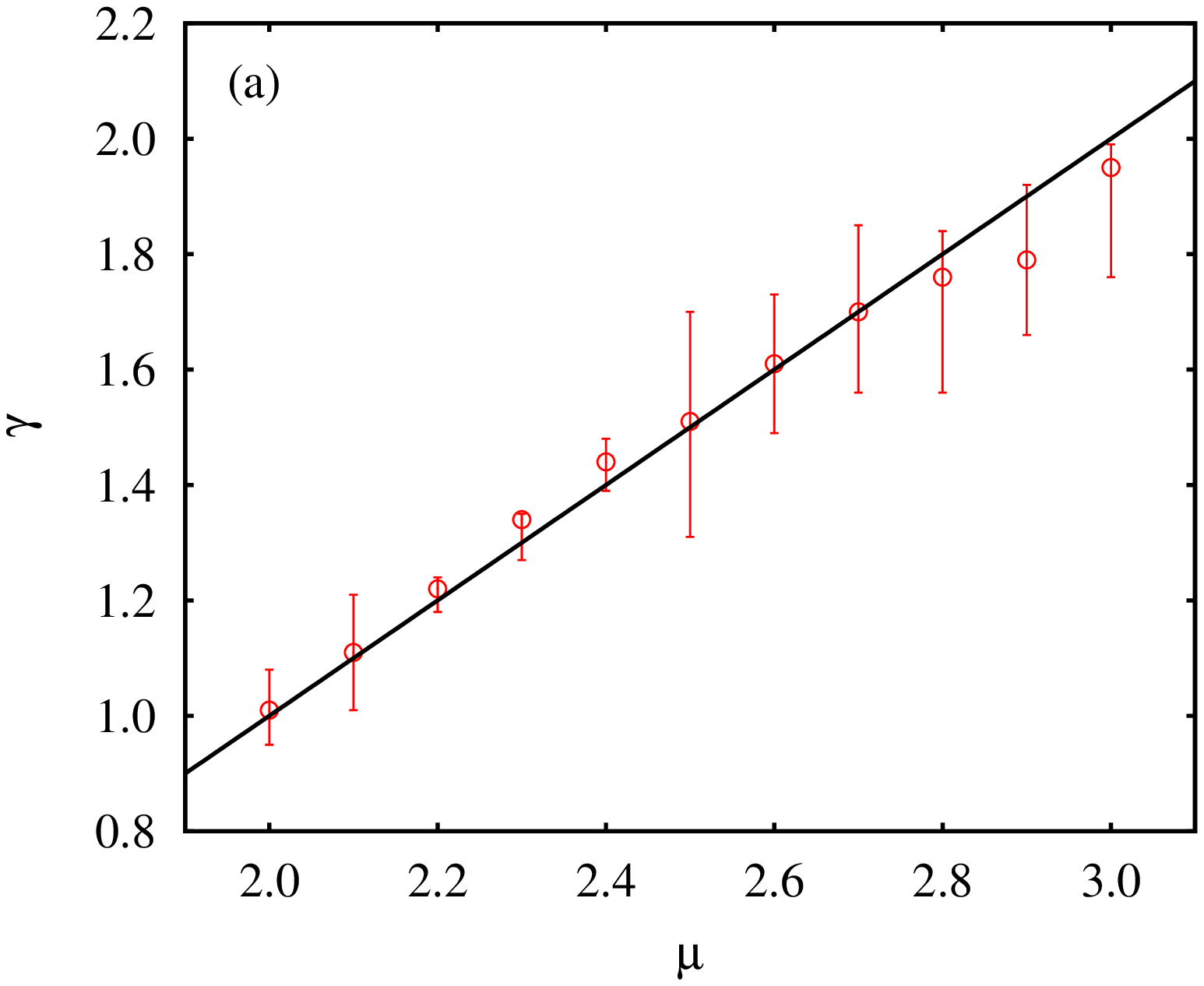}
\includegraphics[scale=0.4]{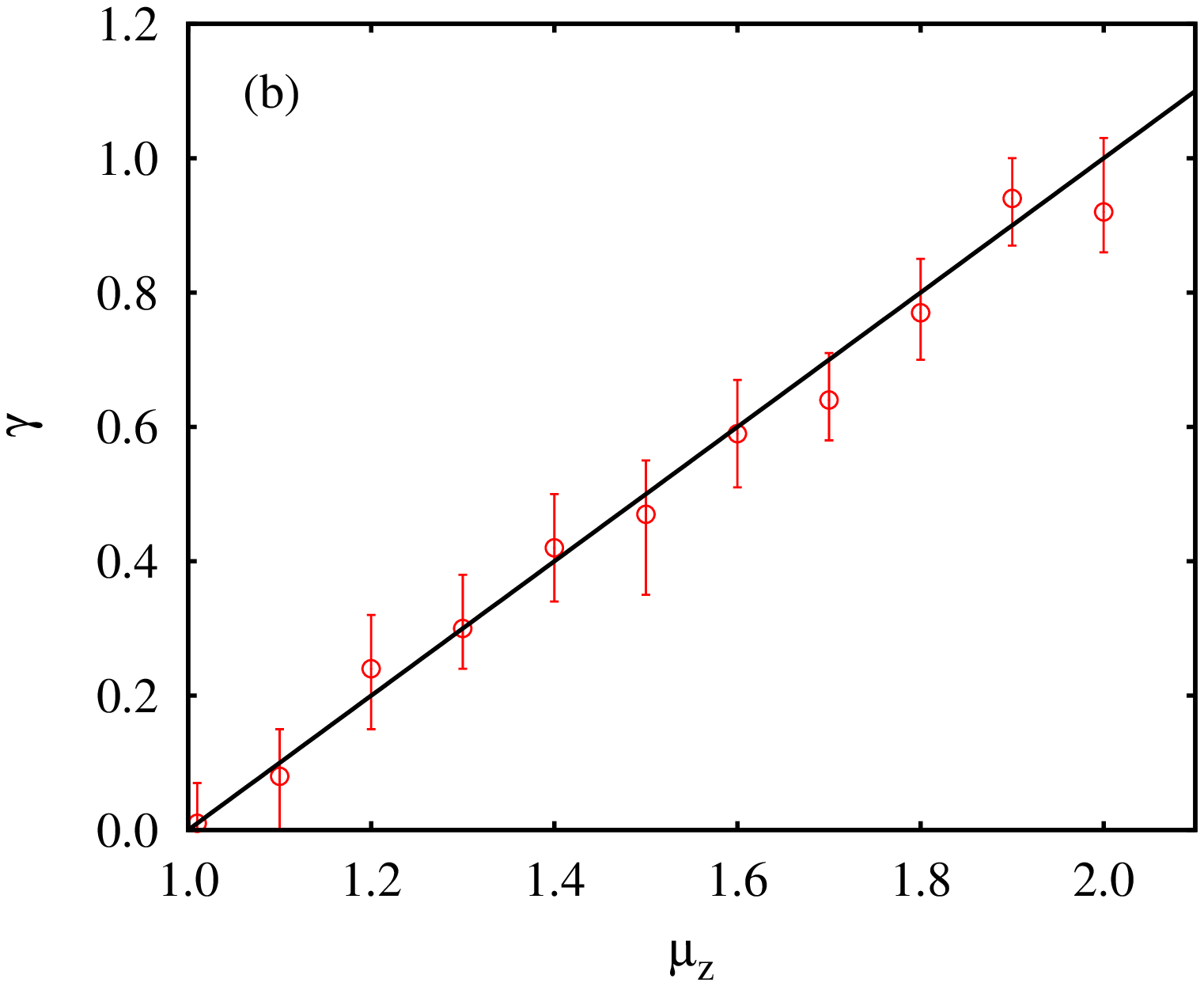}
\caption{(a) The averaged value of $\gamma$ versus $\mu$ obtained via the procedure described in the text for
the equiprobable alphabet $\mathcal A = \{-1,1\}$. (b) The averaged value of $\gamma$ versus $\mu_z$
when considering the equiprobable alphabet $\mathcal A=\{-1,0,1\}$ and $\mu_j=6$. Notice that the relation
$\gamma=\mu-1$ or $\gamma=\mu_z-1$ obtained by comparing $w(t)$ and $p(y)$ is hold in general.
The error bars are calculated via bootstrap resampling method\cite{Efron}.}
\label{fig:params}
\end{figure}

Henceforth we start by considering the coupled velocity model of Zumofen and Klafter\cite{Klafter}
to compare with our sequences. For this case
\begin{equation}
\psi(x,t)=\frac{1}{2}\delta(|x|-t)\,w(t)\,,
\end{equation}
where $w(t)\sim t^{-\gamma-1}$. {Note that in this CTRW model long jumps are
always penalized by long waiting times due to the presence of the $\delta$ function.}
Furthermore, because of the asymptotic behavior of $p(y)\sim y^{-\mu}$, $\gamma$ 
should be related to $\mu$ via \mbox{$\gamma=\mu-1$}. Now, following the approach of Zumofen and Klafter, 
we can write the form of the distribution $p(x,t)$ in the Fourier-Laplace space as
\begin{equation}
p(k,u)= \frac{\Psi(k,u)}{1-\psi(k,u)}\,,
\end{equation}
where
\[
\Psi(x,t)=\frac{1}{2} \delta(|x|-t) \int_t^\infty w(t') dt'
\]
is the probability density to move a distance $x$ in time $t$ in a single event and 
not necessarily stop at $x$. By using $p(k,u)$, we can evaluate the variance
\begin{eqnarray}\label{eq:var}
\sigma^2(t)&=&\mathcal L^{-1}\left[ -\frac{\partial^2}{\partial k^2}p(k,u) 
\Bigl |_{k=0} \right] 
\sim
\begin{cases}
t^2, &  1<\mu<2, \\
t^{4-\mu}, & 2<\mu<3, \\
t, & \mu>3. \\
\end{cases}
\end{eqnarray} 
{Figure \ref{fig:var}b shows the comparison with numerical data for the alphabet $\mathcal A= \{-1,1\}$
and Figure \ref{fig:var2}a makes this for the alphabets $\mathcal A=\{-1,0,1\}$ (with the zero symbol more
probable) and also the larger alphabets $\mathcal A = \{-2,-1,0,1,2\}$ and 
$\mathcal A = \{-20,\dots,-2,-1,0,1,2,\dots,20\}$. Naturally, the agreement is better for the
first case than the others, since it fulfills the requirements of the model. In general, we can see that the
presence of the zero symbol makes the convergence of $\alpha$ to the limiting
regimes slower($\alpha=2$ and $\alpha=1$). 

We consider another CTRW model trying to reproduce the subdiffusive regime. Specifically,
we employ the decoupled version proposed by Montroll\cite{Montroll2} where
$\lambda(x)\sim \exp(-x^2)$ and $w(t)\sim t^{-\gamma-1}$ ($0<\gamma<1$ ).  Following Montroll\cite{Montroll2}
or also Metzler and Klafter\cite{Metzler} we obtain
\begin{equation}\label{eq:var2}
\sigma^2(t)\sim t^{\mu_z-1}\,,
\end{equation}
where again we have used the relation $\gamma=\mu_z-1$. Figure \ref{fig:var2}b confronts the numerical data
with this expression for which we can see a good agreement. 
}

Additionally, we may also obtain the propagator from a small $(k,u)$ expansion for
both previous cases. For the first one $p(k,u)\sim1/(u+c |k|^\mu)$ which for $2<\mu<3$ yields
\begin{equation}\label{pdf}
p(x,t)\sim 
\begin{cases}
t^{-1/\gamma} L_{\gamma}(\xi), & |x|<t,\\
0, &	|x|>t,
\end{cases}
\end{equation}
where $L_\gamma(\xi)$ is the  L\'evy stable distribution and $\xi=c\, x/t^\gamma$ 
is the scaling variable. {For the second one $p(k,u)\sim1/(u+c \,k^2\, u^{\gamma})$
leading to 
\begin{equation}\label{pdf2}
p(x,t)\sim t^{-\gamma/2} H_{1,2}^{2,0} \left[\xi^2|_{(0,1),(1/2,1)}^{(1-\gamma/2, \gamma)}\right]
\end{equation}
where $H_{1,2}^{2,0} \left[\xi^2|_{(0,1),(1/2,1)}^{(1-\gamma/2, \gamma)}\right]$ is Fox $H$ function \cite{Fox} and
$\xi=c x/t^{\gamma/2}$ is the scaling variable.

Figure \ref{fig:histc} shows the comparison for the first case and Figure \ref{fig:histcd} for the second one.
In both cases we can see a good quality data collapse when the scaling is performed. Moreover, these figures 
show that we have found a good agreement between the numerical data and the CTRW models. The fitting
parameter $\gamma$ of each case was obtained by minimizing the difference between the numerical
data and the analytic expressions using the nonlinear least squares method. In all these figures we
have employed the averaged value of $\gamma$ obtained from applying the method for 17 values of $n$ 
chosen logarithmically spaced from $10^3$ to $10^7$. In addition, Figures \ref{fig:params}a and \ref{fig:params}b 
show the dependence of the averaged value $\gamma$ on $\mu$ for both cases, showing that the relation 
$\gamma=\mu-1$ or $\gamma=\mu_z-1$ is consistent with the numerical data.  
}

\section{Summary}
{Summing up, we verified that the symbolic model presented by Buiatti \textit{et al.}\cite{Buiatti}
gives rise to a rich diffusive scenario. Depending on the parameter $\mu$ (or $\mu_z$),
different anomalous diffusive regimes can emerge. Specifically, we have found
subdiffusive, superdiffusive, ballistic, and usual regimes, depending on the model parameters
and also on the trajectories construction. We also investigated the probability distributions of
these processes where non-Gaussians were observed. Our findings support the existence of 
self-similarity in the data, due to the good quality of the data collapse when the scaling is performed. 
In addition, the numerical data were compared with predictions of the CTRW framework finding
a good agreement. We  believe that our empirical findings may help modeling systems
for which power laws are present as well as to motivate other random walk constructions
based on symbolic sequences.}

\acknowledgements
The authors thank CENAPAD-SP (Centro Nacional de Processamento de 
Alto Desempenho - S\~ao Paulo) for the computational support and CAPES/CNPq (Brazilian agencies) 
by financial support. {HVR wishes to acknowledge 
\mbox{Eduardo G. Altmann} for helpful discussions at LAWNP'09.}

\end{document}